\documentclass{aa}
\usepackage{epsfig}

\def\m{$\cal M$~}

\def\hm{$\cal M$$_{HI}$~}
\def\2hm{$\cal M$$_{H2}$~}

\def\ha{H$\alpha$~}
\def\hb{H$\beta$~}

\def\hg{H$\gamma$~}
\def\xha{H$\alpha$}

\def\wha{W(\xha)~}
\def\xwha{W(\xha)}

\def\sma{$\cal M_\odot$}

\begin{document}

\title{Galaxy interactions - poor starburst triggers.\thanks{Based
on observations collected at the European
Southern Observatory, La Silla, Chile.}}

\subtitle{A study of a complete sample of interacting galaxies. III.}

\author{ Nils Bergvall\inst{1}
\and
Eija Laurikainen\inst{2}
\and
Susanne Aalto\inst{3}}

\offprints{N. Bergvall}

\institute{Dept. of Astronomy and Space Physics, Box 515, S-751 20 Uppsala,
Sweden \\
\email{nils.bergvall@astro.uu.se}
\and
Division of Astronomy, Dept. of Physical Sciences, University of Oulu,
FIN-90570 Oulu, Finland \\
\email{eija@sun3.oulu.fi}
\and
Onsala Space Observatory, S-439 92 Onsala, Sweden\\
\email{susanne@oso.chalmers.se}}

\date{Received 0000 / Accepted 0000}

\abstract{
We report on a study of tidally triggered star formation in galaxies based on 
spectroscopic/photometric observations in the optical/near-IR of a magnitude 
limited sample of 59 systems of 
interacting and merging galaxies and a comparison sample of 38 normal isolated  
galaxies.
From a statistical point of view the sample gives us a unique opportunity to 
trace
the effects of tidally induced star formation. In contrast to results from 
previous investigations, our global UBV colours {\it do not} support a 
significant enhancement of starforming activity in the interacting/merging 
galaxies. We also show that, contrary to previous claims, there is no 
significantly increased scatter in the colours of Arp galaxies as compared
to normal galaxies. We {\it do} find support for moderate (a factor of 
$\sim$ 2-3) increase in star formation in the very centres of the interacting 
galaxies of our sample, contributing 
marginally to the total luminosity. The interacting and in particular the 
merging galaxies are characterized by increased far infrared (hereafter FIR) 
luminosities and temperatures that weakly correlate with the central activity. 
The L$_{FIR}$/L$_B$ ratio however, is remarkably similar in the two samples, 
indicating that true starbursts normally are not hiding in the central 
regions of the FIR luminous cases. The gas mass-to-luminosity ratio in 
optical-IR is practically independent of luminosity, lending further support 
to the paucity of true massive starburst galaxies triggered by 
interactions/mergers. We 
estimate the frequency 
of such cases to be of the order of $\sim$ 0.1\% of the galaxies in an 
apparent magnitude limited sample. Our conclusion is that interacting and 
merging galaxies, from the global star formation aspect, generally do not 
differ dramatically from scaled up versions of normal, isolated galaxies. No 
drastic change with redshift is expected. One consequence is that galaxy 
formation probably continued over a long period of time and did not peak at a 
specific redshift.  The effects of massive starbursts, like blowouts caused by 
superwinds and cosmic reionization caused by starburst populations would also 
be less important than what is normally assumed.
}

\authorrunning{Bergvall, N., et al.}

\titlerunning{Star formation in interacting galaxies}

\maketitle

\section{Introduction}
\subsection{Historical background and scientific drivers}

For a long time it has been known that galaxy interactions and mergers are
of fundamental importance for the evolution of galaxies, clusters of
galaxies and the intergalactic medium. This became evident when the first deep 
survey images from HST were analyzed (Abraham et al. \cite{abraham}). But 
already a long time before this, several models focused on the importance of 
mergers for the evolution of structure in the universe and in the interpretation
of the redshift-number density evolution (White \cite{white}, Frenk et al.
\cite{frenk}, Barnes \cite{barnes2}, Rocca-Volmerange \& Guiderdoni
\cite{rocca}, Lacey et al. \cite{lacey}). The analysis of the HST images
allowed a direct morphological study resulting in claims that the merger
frequency increases with redshift (e.g. Patton et al. \cite{patton},
Roche \& Eales \cite{roche}, Le F\`evre et al. \cite{lefevre}). 
These results have immediate implications on our
understanding of evolution of the galaxy luminosity function with redshift
(Mobasher et al. \cite{mobasher}).

In principle, the recent estimates of the
extragalactic background light combined with simulations of
structure evolution in the early universe could be used to obtain interesting 
constraints
on merger rates, star and galaxy formation processes at high redshifts (e.g.
Guiderdoni et al. \cite{bruno}). However, it is important to remember that,
partly as an effect of loosely controlled sample biases in many previous
investigations, there is a lack of quantitative empirical information about
the processes that lead to induced star formation. In particular we want to
know to what extent and under what conditions (relative masses, gas mass
fractions, initial configurations etc.) true starbursts, i.e. corresponding
to gas consumption rates $\ll$ Hubble age, can be tidally triggered and
which effects starbursts have on the intergalactic medium in terms of outflow 
rates and
initial mass function (IMF).

It is normally assumed that the star formation rate (hereafter SFR) in
starbursts is increased with one or two magnitudes to a level at which the gas 
content of the galaxy will be consumed on a time scale short as compared to the
Hubble time. Again, numerical simulations seem to support tidally triggered
nuclear starbursts related to bar formation (Noguchi \cite{noguchi2}, Salo
\cite{salo1}, Barnes \& Hernquist \cite{barnes3}) or encounters between
disks and small satellites (Hernquist \cite{hernquist2}, Mihos \& Hernquist
\cite{mihos}).

Observational indications of starbursts in merging and interacting galaxies
come in many different flavours. The work by Larson and Tinsley
(\cite{larson2}, hereafter LT) has had a major influence on the general 
opinion regarding tidally triggered star formation. Among the other properties 
used to study the effects of interactions are the \ha emission (e.g. 
Kennicutt et al.
\cite{kennicutt2}) and the FIR IRAS emission (e.g. Appleton \&
Struck-Marcell \cite{appleton}, Kennicutt et al. \cite{kennicutt2}, Bushouse
et al. \cite{bushouse1}, Kennicutt \cite{kennicutt3}, Sanders \& Mirabel
\cite{sanders}). Almost all find strong support of tidally induced starbursts. 
The effect is found 
to 
be strongest in the nucleus but the star formation rate is also enhanced in 
the disk (as a counterexample see e.g. Hummel \cite{hummel}). Extended emission 
in the radio continuum in dusty galaxies as well 
as direct evidences for global outflows of gas (e.g. Heckman et al. 
\cite{heckman}) have also been put forward as evidence for a dramatically 
increased supernova activity from a starburst region. Although some of these 
observations seem compelling, there is no unique way to interpret them. A 
widely accepted idea is that luminous IRAS galaxies that show no signs of 
an active nucleus are strong starbursts. Such a generalization is doubtful 
however, since there are
alternative explanations to the observed FIR fluxes. For example Thronson
et al. (\cite{thronson}) presents a scenario where the fragmentation  and
disruption of dust clouds may lead to an increase in the efficiency of the 
dust heating without the addition of new heating sources via a starburst. 
The warmer
dust will result in an increase in the integrated FIR flux.  There is, of
course, compelling evidence for interacting galaxies having an increase 
in star formation activity but the important questions are - what
is the level of increase and how frequently do these events occur?

\begin{figure}
\epsfig{file=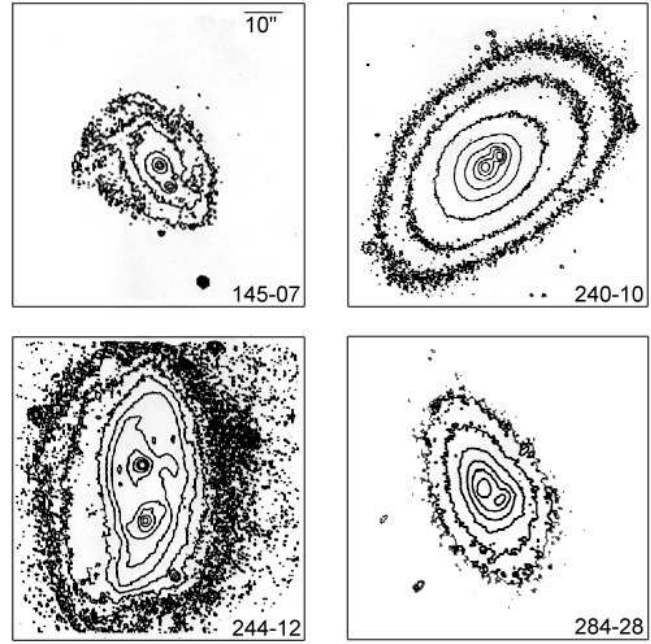, width=\hsize}
%	\resizebox{\hsize}{!}{\includegraphics{H3242f1.ps}}
      \caption[]{Johnson H images of double nuclei in four merger candidates. 
The
ESO numbers are indicated. ESO/MPI 2.2-m telescope.}
\label{nuclei}
   \end{figure}

One of the most interesting problems in this context is the origin of
elliptical galaxies and massive spirals. Do they essentially form from a
single gas cloud at high redshift (the monolithic scenario) or as a result
of a series of mergers (the hierarchical scenario)? The hierarchical 
scenario,
first suggested by Toomre (\cite{toomre2}) has been defended by e.g.
Schweizer (\cite{schweizer3}) and Kormendy (\cite{kormendy}). Strong 
support of recent mergers and cannibalism in samples of nearby galaxies 
is found in
morphological distortions like tails, shells (Schweizer \cite{schweizer1},
Quinn \cite{quinn}, Schweizer and Seitzer \cite{schweizer3}), boxy isophotes
and double nuclei in bright ellipticals (e.g. Schweizer \cite{schweizer5}).
A large fraction of our merger candidates also have double nuclei. As an
example figure \ref{nuclei} shows images of four merger candidates of various 
morphological types obtained in the H band (minimizing the risk of 
misinterpreting the double structure due to extinction effects). 
Compelling 
support of hierarchical galaxy formation also comes from observations at 
intermediate redshifts (Franceschini et al. \cite{franceschini}).

\begin{table*}

\caption[ ]{Comparison between our photometry JB (Johansson \& Bergvall
\cite{johansson3}) and other published data. $\Delta$V, $\Delta$(B-V) and
$\Delta$(U-B) are the differences between our data and the other 
published data.}
\begin{flushleft}
\begin{tabular} {lllllllll}
\noalign{\smallskip}
\hline
\noalign{\smallskip}

ESO-nr      &    D"  &  V  & $\Delta$V &  B-V & $\Delta$(B-V) &  U-B & 
$\Delta$(U-B)& Reference \\
\hline
\noalign{\medskip}

026- G04     & 31.2 & 12.78 && 1.22  && 0.56 && JB \\
                & 31.2 &   12.92 & 0.14 &  1.11 & 0.11 &  0.62 & -0.06	
&Lauberts  \cite{lauberts} \\
109-IG22 W  &  31.2 & 12.86  & & 0.96 & & 0.56 & & JB \\
            &  31. &  12.93 & -0.07 & 0.99 &-0.03 & 0.51 & 0.05&West et al. 
\cite{west} \\
110- G22 W &   60.8 & 13.96 & & 0.68 & &-0.18 && JB  \\
             &   62. &  13.82 & 0.14 & 0.73 & -0.05 &&	&Peterson 
\cite{peterson} \\
110- G23 E &   60.8 & 13.29 &&  0.85 &&  0.19 && JB  \\
           &   62. &  13.29 & 0.00 &  0.84 & 0.01 &  &   &Peterson 
\cite{peterson} \\
145-IG07   &   31.2 & 14.58 &&  0.93 &&  0.82 && JB  \\
           &   25. &  14.55 &&  0.95&&	&&	Peterson \cite{peterson} \\
           &   36. &  14.27 &&  0.85 && &&		" \\
           &   31.  & 14.39 & 0.19 &  0.89 & 0.04 && &		", interpolated 
data \\
148-IG10   &   31.2 & 14.15 &&  0.49 && -0.29 && JB \\
           &   25. &  14.50 &&  0.50&&&&		Peterson \cite{peterson} 
\\
            &  36. &  13.97 &&  0.47&&&&		" \\
            &  31.&   14.20 & -0.05 &  0.48 & 0.01 &&&		", interpolated 
data \\
157- G22    &  31.2 & 11.36&&0.96 &&  0.59 && JB \\
            & 31.5 & 11.36 & 0.00 &  0.96  &0.00&&& Chincarini et al.
\cite{chincarini} \\
            &  30. &  11.42 & -0.06&  0.90& 0.06 & 0.64 &-0.05   & Sandage and 
Visvanathan
\cite{sandage2} \\
193- G19 N  &  31.2 & 14.38 &&  0.87&&   0.90 && JB \\
            &  25.  & 14.55  && 0.99&&&&Peterson \cite{peterson} \\
            &  36. &  14.30  && 0.90&&&&		" \\
            &  31. &  14.41 & -0.03 &  0.94& -0.07 &&&		", interpolated 
data \\
200-IG31 N  &  43.4 & 13.69 &&  0.84&&   0.38 && JB   \\
            &  43.4 & 13.75 & -0.06 &  0.76 & 0.08 &&  & Chincarini et al. 
\cite{chincarini} \\
233- G21    &  31.2 & 12.61 &&  1.00 &&  0.52 && JB \\
             & 60.8&12.16 &&  0.98 & & 0.49&&	JB	 \\
            &  30. &  12.57 & 0.04  & 0.92 & 0.08 & 0.51& 0.01&	Sandage and 
Visvanathan
\cite{sandage2} \\
             & 52.  & 12.28& -0.12 &  0.94 &0.04 &  0.48 & 0.01&Shobbrook 
\cite{shobbrook} \\
236- G01     & 86.6 & 11.01 &&  0.80 &&  0.00 && JB \\
            &  89.  & 10.95 & 0.06 &  1.06 & -0.26 & 0.47 & -0.47 &Alcaino 
\cite{alcaino} \\
249-IG31    &  86.6 & 12.34 &&  0.36 && -0.31&&		JB, Johnson filters \\
            &  86.6 & 12.28  && 0.35 && -0.34&& "		\\
            &  87.  & 12.11 &0.13  & 0.38 & -0.02 & -0.29 & -0.02   &  Bergvall 
et al.
\cite{bergvall5} \\
284- G28 S  &  31.2 & 12.67 &&  0.97&&   0.42 & & JB \\
               &&     12.64& &  0.98 &&  0.38&&		Lauberts  
\cite{lauberts} \\
		&&    12.69& -0.02 &   1.01 & -0.04 &  0.36 & 0.06 &Sandage \& 
Visvanathan \cite{sandage2} \\
286-IG19     & 22.9 & 14.70 &&  0.61 && -0.06  && JB \\
            &  22.  & 14.79 & -0.09 &  0.53 & 0.08 &-0.02& -0.04 &Bergvall et 
al. 
\cite{bergvall5} \\
287- G17    &  43.4 & 13.20 &&  1.01&&   0.48 && JB \\
            &  43. &  13.29 & -0.09 &  0.95& 0.06 &&&Sadler \cite{sadler} \\
299- G07     & 86.6 & 11.65 &&  0.81 &&  0.14&& JB \\
            &  81. &  11.68 & -0.03 &  0.79 & 0.02 &&&Peterson \cite{peterson2} 
\\
            &  91. &  11.56 & 0.09 & 0.78 & 0.03  & 0.10 & 0.04 &Griersmith 
\cite{griersmith} 
\\
             & 91.  & 11.53 & 0.12 &  0.79 & 0.02 &0.11 & 0.03 &Griersmith 
\cite{griersmith} \\
\noalign{\bigskip}
Mean $\Delta$, m.e. &&& 0.01, 0.11 && 0.02, 0.06 && 0.01, 0.04 & * \\
\noalign{\smallskip}
Median $\Delta$  &&& 0.00 && 0.02 && 0.01 &  \\
\noalign{\smallskip}
Median $|\Delta|$ &&& 0.06 && 0.04 && 0.04 & \\
\noalign{\smallskip}
\hline
*) ESO 236-G01 excluded
\end{tabular} \\
\end{flushleft}
\end{table*}
\smallskip

\begin{table}

\caption[ ]{Heliocentric velocities and formal mean errors of the galaxies
in
the
IG sample. n is the number of absorption (a) and emission (e) lines used in
the
determination of the velocities.}
\begin{flushleft}
\begin{tabular} {llll}
\noalign{\smallskip}
\hline
\noalign{\smallskip}
ESO-nr      &        v  &  $\sigma _v$ &   n\\
&   km s$^{-1} $  &  km s$^{-1}$ &   \\
\hline
\noalign{\medskip}

073-IG32 S  &    5305 &103 & 4a,0e \\
079-IG13 E  &   11557 &107 & 3a,0e \\

079-IG13 W  &   11526 & 12 & 3a,1e \\
079- G16    &    5661 & 11 & 0a,9e \\
080-IG02 W  &    7509 & 38 & 0a,4e \\
080-IG02 E  &    7301 &   - & 1a,0e \\
085-IG05    &    6160 & 49 & 0a,3e \\
105- G26 W  &   10911 &131 & 3a,1e \\
108-IG18 W  &    8010 & 49 & 2a,1e \\
108-IG18 E  &    7908 & 17 & 0a,7e \\
108-IG21    &    3423 & 69 & 3a,5e \\
109-IG22 E  &    3291 & 64 & 6a,0e \\
109-IG22 W  &    3320 & 19 & 4a,0e \\
110- G22 W  &    9755 & 41 & 6a,4e \\
110- G23 E  &   10189 & 19 & 3a,0e \\
112- G08A   &   10167 & 16 & 5a,0e \\
117- G16    &   10612 &   - & 0a,1e \\
143- G04    &   15218 & 30 & 3a,2e \\
145-IG07    &    8540 & 27 & 3a,0e \\
145-IG21 N  &   19869 & 18 & 2a,2e \\
145-IG21 S  &   20145 & 36 & 1a,8e \\
148-IG10    &    3201 & 53 & 1a,4e \\
151-IG36 W  &    3233 &119 & 1a,4e \\
151-IG36 E  &    3349 & 37 &11a,7e \\
157-IG05    &    1141 & 48 & 2a,2e \\
157-IG50 W  &    3760 & 37 & 0a,9e \\
186- G29 N  &    2708 & 49 & 2a,1e \\
187-IG13 S  &   12838 &   - & 0a,1e \\
187-IG13 N  &   13654 & 39 & 1a,1e \\
188-IG18 W  &    4922 & 54 & 3a,5e \\
188-IG18 E  &    4761 & 27 & 3a,6e \\
193- G19 N  &   10291 & 27 & 4a,2e \\
199- G01    &    9046 &141 & 2a,2e \\
200-IG31 N  &   12462 &   - & 0a,1e \\
200-IG31 S  &   11422 &   - & 0a,1e \\
205- G01    &     546 & 17 & 0a,9e \\
235-IG23 S  &    6678 & 98 & 4a,0e \\
235-IG23 N  &    6725 & 62 & 3a,1e \\
240- G10    &    3474 &   - & 1a,0e \\
240- G10 NW &    3315 &100 & 2a,0e \\
240- G01 W  &   15406 & 42 & 9a,0e \\
243- G15 N  &    7014 & 46 & 5a,1e \\
244- G12 N  &    6274 &  8 & 2a,8e \\
244- G12 S  &    6339 & 18 & 3a,6e \\
244- G17 W  &    6910 & 40 & 3a,2e \\
244- G17 E  &    6954 & 48 & 2a,2e \\
244-IG30    &    7193 & 60 & 3a,3e \\
244- G46 E  &    5908 & 99 & 3a,1e \\
245- G10    &    6070 &   - & 1a,0e \\
249-IG31    &     329 & 68 & 0a,5e \\
284- G28 S  &    3049 &   - & 1a,0e \\
\noalign{\smallskip}
\hline
\end{tabular} \\
\end{flushleft}
\end{table}

\begin{table}

\caption[ ]{Heliocentric velocities and formal mean errors of the galaxies
in
the IG sample, continued.}
\begin{flushleft}
\begin{tabular} {llll}
\noalign{\smallskip}
\hline
\noalign{\smallskip}
ESO-nr      &        v  &  $\sigma _v$ &   n\\
&   km s$^{-1} $  &  km s$^{-1}$ &   \\
\hline
\noalign{\medskip}

284-IG41 N  &    4932 &119 & 4a,0e \\
284-IG41 S  &    5165 & 41 & 8a,1e \\
284-IG45    &    5324 &201 & 4a,0e \\
284-IG48    &    5152 &128 & 2a,2e \\
285- G04    &   15979 &155 & 3a,0e \\
285- G13    &    3115 & 65 & 4a,0e \\
285-IG35    &    9002 & 73 & 3a,1e \\
286-IG19    &   12975 & 57 & 3a,4e \\
287- G40    &    8976 & 96 & 1a,1e \\
288- G32 E  &    6058 &   - & 1a,0e \\
288- G32 W  &    7325 & 90 & 3a,6e \\
290- G45    &    2378 &137 & 5a,7e \\
293- G22 N  &    6681 & 66 & 2a,2e \\
297- G11 W  &    4822 &111 & 2a,3e \\
297- G12 E  &    4848 &146 & 2a,4e \\
299-IG01 N  &    5366 &153 & 5a,0e \\
299-IG01 S  &    5773 &148 & 3a,4e \\
303- G17 W  &    3834 & 98 & 2a,0e \\
306- G12 S  &   10994 &   - & 1a,0e \\
340- G29    &    9283 & 14 & 2a,2e \\
341-IG04    &    6225 & 72 & 6a,0e \\
342-IG13 N  &    2754 & 19 & 0a,5e \\
344- G13 S  &   10639 & 63 & 5a,5e \\
\noalign{\smallskip}
\hline
\end{tabular} \\
\end{flushleft}
\end{table}

\begin{table}

\caption[ ]{Heliocentric velocities and mean errors of the galaxies in the
NIG
sample.}
\begin{flushleft}
\begin{tabular} {llll}
\noalign{\smallskip}
\hline
\noalign{\smallskip}
ESO-nr      &        v  &  $\sigma _v$ & n \\
&   km s$^{-1}$  &  km s$^{-1}$ &   \\
\hline
\noalign{\medskip}

015- G05    &    4919 & 30 & 6a,1e \\
026- G04    &    2921 & 24 & 7a,1e \\
027- G14    &    4591 & 21 & 0a,5e \\
047- G19    &    3135 & 52 & 5a,0e \\
048- G25    &   11262 & 33 & 0a,2e \\
052- G16    &    8079 & 23 & 5a,0e \\
074- G26    &    3246 & 51 & 2a,2e \\
105- G12    &    4127 & 44 & 5a,0e \\
106- G08    &    3248 & 29 & 2a,1e \\
109- G15    &    3552 & 34 & 6a,0e \\
114- G16    &    7171 & 82 & 3a,0e \\
151- G43    &    5051 & 73 & 2a,2e \\
153- G01    &    6885 & 26 & 5a,0e \\
153- G33    &    5772 & 24 & 4a,2e \\
157- G22    &     958 & 36 & 8a,0e \\
193- G09    &    6156 &   - & 1a,0e \\
197- G18    &    5893 &128 & 6a,0e \\
201- G12    &    1045 &   - & 0a,1e \\
200- G36    &    1064 & 41 & 3a,3e \\
233- G21    &    3140 & 27 & 8a,0e \\
236- G01    &    2450 & 55 & 6a,0e \\
240- G12    &    1819 & 14 & 0a,6e \\
242- G05    &    6004 & 29 & 7a,0e \\
285- G13    &    3115 & 65 & 4a,0e \\
287- G17    &    5385 & 15 & 6a,0e \\

287- G21    &    6232 &  9 & 1a,1e \\
293- G04    &    1806 & 24 & 1a,4e \\
298- G27    &    5189 &  9 & 2a,0e \\
299- G07    &    1743 & 20 & 1a,5e \\
299- G20    &    1686 & 36 & 1a,8e \\
303- G14    &    6208 & 27 & 4a,5e \\
304- G19    &    6478 & 21 &10a,0e \\
340- G07    &    6100 & 28 & 3a,3e \\
341- G32    &    2792 &   - & 0a,1e \\
345- G49    &    2475 & 26 & 1a,2e \\
\noalign{\smallskip}
\hline
\end{tabular} \\
\end{flushleft}
\end{table}

It has long been suspected that strong interactions and mergers may trigger
nuclear activity. Indeed, most of the quasar host galaxies show distorted
morphologies, reminiscent of the aftermaths of mergers (McLeod \& Rieke
\cite{mcleod1}, \cite{mcleod2}, Bachall et al. \cite{bachall}). It is not
immediately clear however, that the reverse situation is true.
Even though Dahari (\cite{dahari1}) first suggested that Seyferts 
have more frequently companions than the non-active galaxies, this
probably is not the case. Laurikainen \& Salo (\cite{laurik2}) have reviewed 
that in that kind of comparisons the apparent controversies between
different authors can be
largely explained by selection effects. Also, Barton et al. (\cite{barton})
shows no elevation in counts of Seyferts and active galaxies among galaxies 
in pairs.

The major goal with the present work is to try to quantify the effects of
interaction on star formation and nuclear activity in a unique way. We will
compare two samples of galaxies. One sample contains isolated pairs of
interacting galaxies and merger candidates and the other consists of isolated 
single
galaxies.  

   \begin{figure}
	\epsfig{file=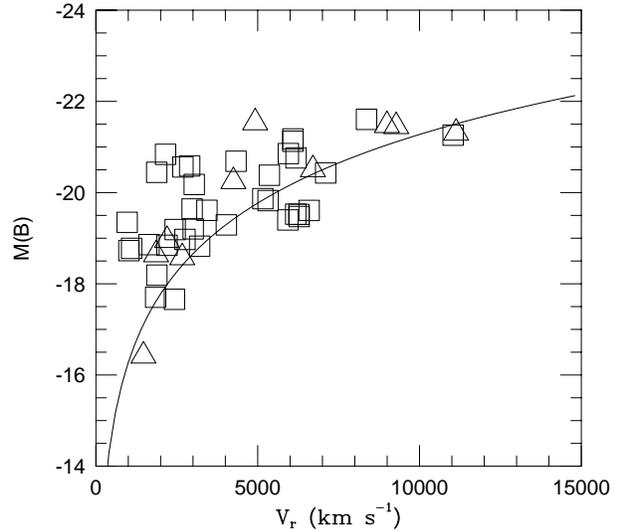, width=8cm}
      \caption[]{The distribution of the comparison galaxies in radial velocity
and absolute magnitude, M$_B$. M$_B$ is based on total magnitudes obtained
from LEDA, corrected for galactic extinction. The triangles are galaxies in 
the additional sample discussed in Sect. 3.4. The 
solid line marks the 
m$_B$=14.5 magnitude limit.}
         \label{rvel}
   \end{figure}
\section{The samples and data extraction}

Several samples of interacting and merging galaxies have been used in
previous studies in order to tackle the issues discussed above. However,
few are based on selection criteria that open a possibility to relate
the results to the galaxy population in general in an unbiased way.
The results from these studies consequently are contradictive.

Here we discuss a spectroscopic/photometric study of a magnitude-limited
sample of interacting and merging galaxies and of isolated galaxies for
comparison. The samples are presented by Johansson \& Bergvall
(\cite{johansson3}, henceforth JB). It is based on a catalogue by Bergvall
(\cite{bergvall3}) containing about 420 interacting galaxies
and merger candidates and a comparison sample of about 320 isolated galaxies 
from the
ESO/Uppsala Quick Blue Survey (\cite{holmberg}, \cite{lauberts2}). In Bergvall's 
sample a
merger is defined as an isolated galaxy (no obvious companion within a
projected distance of 6 diameters) with a strongly distorted morphology.
 Based on the statistics of the large sample of galaxies Bergvall
(\cite{bergvall4}) argues that a substantial part of
these cases actually may be interacting pairs with the separation vector
closely aligned with the line of sight, causing one component to hide the
other from view. These hidden pairs are appearing as mergers in our
sample. If it is assumed that the major/minor axes ratios of the cases
classified as mergers are the same as that of the major components in
strongly interacting pairs, almost all such cases may be major components of
close pairs that are hiding the companions.
In the following, for simplicity, we will sometimes use the word merger 
although it would be more proper to say merger candidate since we cannot claim 
to have sufficient information to prove that all merger-like cases are true 
mergers.

From the catalogue we selected {\it 59 pairs and clear cases of
mergers} (hereafter IG) complete down to about m$_B$ = 14.5$\pm$0.3 mag. 
for a spectroscopic/photometric study. The non-interacting comparison sample
contained 38 isolated galaxies (hereafter NIG). These were defined as 
galaxies having no neighbours (with a magnitude difference $\leq $ 2 
magnitudes) closer than 6 diameters and having not more than 2 neighbours 
within 16 diameters. The limiting magnitude is the same as for the IG 
sample but the sample is not required to be complete (see below the discussions 
regarding the morphological selection and the luminosity function). In Bergvall 
and Johansson (\cite{bergvall1}, 
henceforth BJ) we published the images, colour maps, spectra and energy 
continuum distributions of the IGs. Here some of the remaining optical data and 
an 
analysis of the global and nuclear star formation properties of the 
two samples are presented.

Since the following discussion will result in conclusions that deviate from
those of many previous investigations of similar samples it is important to
compare our criterion of what we regard as a galaxy with distorted 
morphology with that of others. A suitable comparison sample is the 
Arp-Madore sample of southern peculiar galaxies and associations (Arp \& 
Madore \cite{arp2}),
obtained from a region of the sky including our sample. We find that 84\% of
our galaxies are included in the Arp-Madore catalogue. Among the remaining 11
cases, a few are wide pairs, others are pairs with low luminosity companions 
and some may be dwarf galaxy mergers.

In the following discussion it is also important to be aware of the 
possible selection effects occurring when we compare the two samples. 
The basic idea of this investigation is to find out what kind of changes 
appear in the physical conditions when galaxies are subject to 
interactions and mergers. If, as is often argued, the mean luminosity 
increases and the spectral distribution changes, it will influence the 
luminosity function. If one wants to
compare non-interacting with interacting galaxies it would therefore be
wrong to demand that the luminosity functions of the two samples should 
agree.
Should this demand be applied it would e.g. result in a bias towards higher
masses of the sample of non-interacting galaxies if interactions lead to 
an increasing star formation rate. If IGs normally live in environments 
where mergers are more frequent than in normal environments, this will 
also cause a shift of the luminosity function causing ambiguities. 

The method we will use here is to select a comparison sample of isolated 
galaxies with a distribution of morphological types and luminosities that 
agrees reasonably well with that of the galaxies {\it originally} involved 
in the interaction/merging. All efforts to find out the differences 
between interacting and normal galaxies by comparing two different samples 
have their problems, since we a priori do not know the differences in 
evolutionary history of IGs and NIGs, but we think that our approach will 
lead to a result that will be concise and straightforward to interpret. 
It is impossible to derive the original properties of the galaxies of the 
IG sample but we can do our best to make a classification of the IGs and 
then make an appropriate selection of galaxies in the comparison sample
that have morphologies that we think reasonably well represent those of
the galaxies in the IG sample. In the original
classification system that Bergvall used, the galaxies were morphologically
classified according to: E: ellipticals or S0s; S-: early spiral
galaxies; S+: late type spirals and irregulars; S: spiral galaxies
difficult to classify; D: dwarfs; C: compact galaxies and
G: unclassifiable. In this system, the two different samples under study
here have the
following distributions of morphological types. IGs: 27\% E, 25\% S-,
27\% S+, 10\% S and 10\% G. NIGs: 31\% E, 36\% S-, 31\% S+ and 1\% S.
If one only compares the relative numbers of the galaxies with reliable
classification,  the match between the morphological types is good
(34/31/34 \% and 31/36/32 \% respectively). With few
exceptions (Kennicutt et al. \cite{kennicutt2}) there is no other similar
investigation that has taken into account the effects of morphological
selection. Therefore they run a severe risk of biasing when comparing
samples of interacting and noninteracting galaxies.

   \begin{figure}
%\sidecaption
%\includegraphics[width=9cm]{H3242f3.ps}
\epsfig{file=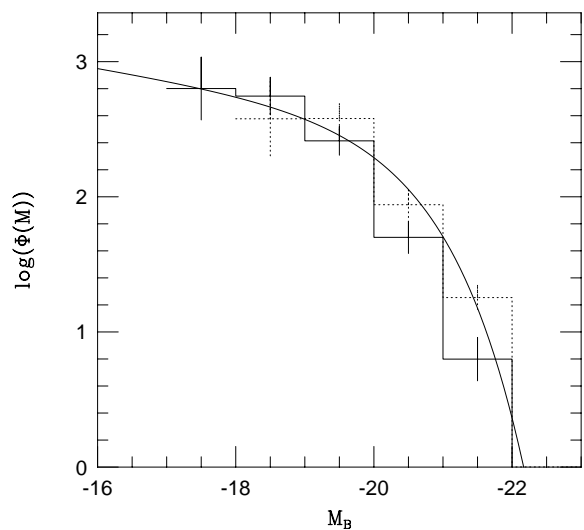, width=8cm}	
%\resizebox{\hsize}{12cm}{\includegraphics{H3242f3.ps}}
\caption{The luminosity function (LF) of the comparison sample 
(solid histogram).
The vertical bars are the errors based on Poisson statistics. The hatched 
line
illustrates how the LF would change if the galaxy population was influenced 
by
mergers according to the observational results by Le F\`evre et al. 
(\cite{lefevre}).
The solid line is the LF of field galaxies according to Ramella et al. 
(\cite{ramella}).}
         \label{lumf}
   \end{figure}

One may worry that the morphological selection may have introduced a bias 
in the LF of the NIG sample. This should not be a problem since we throw out 
only about 15\% of the galaxies in the total sample of NIGs down to the limiting 
magnitude from Bergvalls original catalogue. Still, it seems to be a fact that 
the IG sample contains a larger proportion of luminous galaxies, despite that 
the limiting magnitude of the two samples is the same. Could there be other 
biases involved in the selection? We will make two tests that will demonstrate 
that this is not a problem.

Fig. \ref{rvel} shows the distribution of total apparent B magnitudes of the 
NIG sample plotted against radial velocity. The magnitudes were obtained from
LEDA (the Lyon-Meudon extragalactic database, through the CISM of the Lyon
Claude-Bernard University) and have been corrected for galactic extinction
(Burstein \& Heiles \cite{burstein}). The original magnitude
limit (m$_B$= 14.5$\pm$0.3 mag.) has been indicated and the agreement
is good. A few additional
isolated galaxies, used later in this paper, have also been included
to increase the contrast at the high luminosity tail. As we can see
from the diagram, and as will be confirmed in the next section, there does 
not seem to be any problem with the completeness
at the high luminosity tail since it seems to be reached already at rather
low redshifts. Fig. \ref{lumf} shows the LF of the NIG sample (not including 
the 'extra' isolated galaxies), as obtained after
volume corrections of the numbers at each magnitude bin, assuming constant
space density of the galaxies. A comparison with the LF of field galaxies
(Ramella et al. \cite{ramella}) shows  that there is a relative 
underrepresentation of
bright galaxies in our NIG sample. The evolution of isolated galaxies
differ however from normal field galaxies in the sense that the field
galaxies in the mean have experienced more merger events. If we apply
a compensation for this fact by adding 0.5$^m$ to our NIG sample,
corresponding to the increase in luminosity of a typical L$_*$ galaxy
due to mergers between z=1 and 0 (Le F\`evre et al. \cite{lefevre}), we 
obtain full agreement.

   \begin{figure}
	\epsfig{file=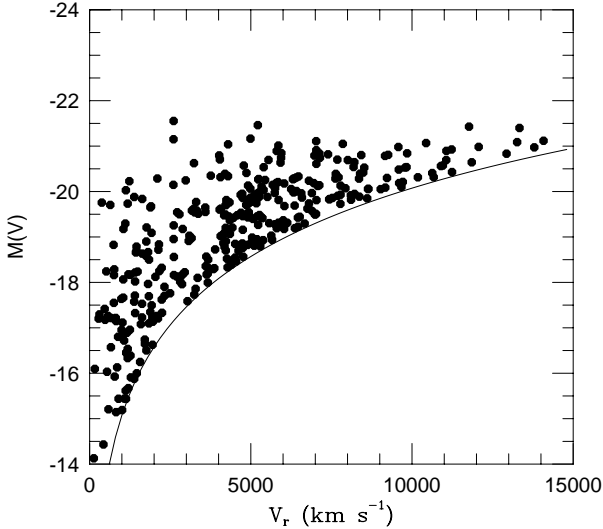, width=8cm}
%\resizebox{\hsize}{!}{\includegraphics{H3242f4.ps}}
      \caption[]{The distribution of the isolated galaxies of the Karachentseva 
(\cite{kara1}, \cite{kara2}) sample in the radial velocity-M$_V$ plane. The 
solid line marks the 
m$_V$=15.7 magnitude limit.}
\label{kara}
   \end{figure}

\begin{figure}
\epsfig{file=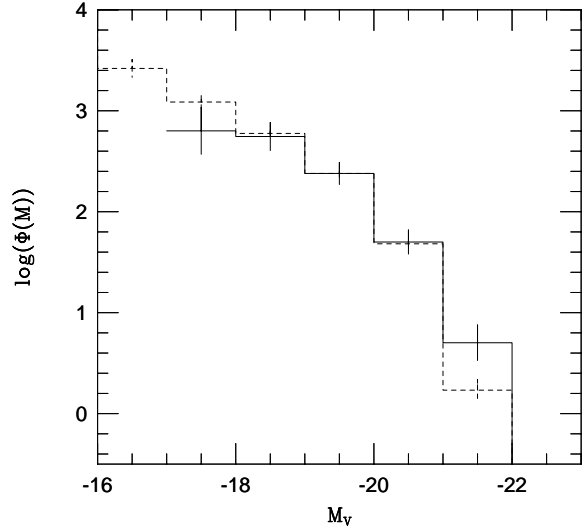, width=8cm}
%	\resizebox{\hsize}{!}{\includegraphics{H3242f5.ps}}
      \caption[]{The L(V) luminosity functions our sample (solid line) and 
Karachentsevas sample of isolated galaxies (hatched line). The V magnitudes of 
our sample are based on the B magnitudes obtained from LEDA combined with our 
B-V colours and a correction of 0.2 magnitudes to compensated for the difference 
in galactic extinction between our sample and Karachentsevas sample and 0.5 
magnitudes which is the approximate difference in zeropoint between the LEDA and 
the Karachentseva total V magnitude scales. The two distributions have been 
adjusted to agree at the central magnitude bin.}
\label{karalf}
   \end{figure}

Now we will discuss a second approach that more directly supports our claim 
of completeness. Fig. \ref{kara} shows the distribution in the 
redshift-magnitude diagram of isolated galaxies (isolation class:0) obtained 
from the list of isolated galaxies by Karachentseva (Karachentseva \cite{kara1}, 
Karachentseva et al. \cite{kara2}). The limiting magnitude is m$_V$=15.7. This 
is deeper than the limiting magnitude of our sample. Thus is is possible to 
compare the two samples, one definitely reaching completeness at the high 
luminosity tail, as is seen from 
the figure, and the second our comparison sample. As 
the diagram shows, there is no doubt that we reach the high luminosity tail of 
the LF at v $\approx$ 5000 km s$^{-1}$, i.e. below the velocity limit of our 
sample. To quantify this fact we show in Fig. \ref{karalf} a comparison between 
the LF of our sample and that of Karachetseva. The V magnitudes of our sample 
have been calculated from the LEDA B magnitude data and transferred to V, using 
our B-V data. A correction of 0.2 magnitudes was then applied on our V 
magnitudes to approximately correct for the difference in galactic extinction 
between our sample and Karachentsevas. A zeropoint correction term of 0.5 
magnitudes was also added to our data. This correction is based on a comparison 
between LEDA data of a few of the brighter galaxies in the Karachentseva sample 
but the exact value to within a fraction of a magnitude is not important for 
the result of the comparison. As we see from the diagram the LFs agree well 
within the statistical noise exept in the fainter end where the dwarfs in our 
sample normally were given a lower weight in the selection since they do not 
occur frequently in the IG sample. We also note that there is no lack of 
luminous galaxies in our sample as one might have suspected. Instead there is a 
small overrepresentation 
at higher luminosities that probably is due to the morphological 
selection. But the differences are small and one has to keep in mind that it is 
not completely adequate to compare the samples since one was selected on V 
magnitudes while our galaxies were selected on B magnitudes. Anyway, these 
discussions show that the LF of our sample is representative of a sample of 
isolated galaxies at the same time as it has a relevant morphological 
distribution in comparison with the IGs.

In Tables 1-5 the results of the spectroscopy of the central 1.5"x3" of the 
sample galaxies are presented. The fact that the redshift
distribution is different for the IG and NIG will result in a different
mean absolute size of the sampled central region. However, based on
estimates by Carter et al. (\cite{carter}) , the effect on the equivalent widths
will be insignificant in our case. Tables 2-4 contain the radial velocity 
data and Tables 5-6 the emission line flux densities. The emission
line data were corrected for
atmospheric and Galactic extinction as described in BJ. The line 
intensities
were measured by fitting a gaussian profile to the observed line profile.
Blended lines were deblended by assuming the wavelength shift between the
lines to be known and then applying an optimized double gaussian fit. The
underlying continuum was approximated with a straight line. Each line
measurement was tagged with a weight that was calculated from the noise
statistics in the gaussian fit. The
weight depends on the noise in a manner that was derived from measurements
of synthetic spectral lines that were degraded with different amounts of
noise.
The velocities were then calculated from the weighted mean of the absorption
or
the emission lines separately. Some emission
lines,
like [OIII]$\lambda$4959, are sometimes strongly hampered by Telluric
atmospheric
emission lines and are therefore not included in the tables. Normally the
error
in the line intensities are estimated to be about 10\% for the brighter
lines
and about 20\% for the weakest ones tabulated.

Throughout this paper we will assume a Hubble parameter of
H$_0$=70 km s$^{-1}$ Mpc $^{-1}$.

\section{Star formation properties}
\subsection{Introduction}

The idea that interacting galaxies experience an enhanced star formation
activity as compared to noninteracting galaxies was permanently established
by LT in their analysis of the broadband UBV
colours. They compared two different samples
at a Galactic latitude b $\geq$ 20$^\circ$ - the noninteracting galaxies
taken from the Hubble Atlas of Galaxies by Sandage
(\cite{sandage}; hereafter Hubble galaxies)  and interacting galaxies
from Arp's Atlas of Peculiar Galaxies (\cite{arp}; hereafter Arp galaxies).
LT found that 1) the scatter in the U-B/B-V diagram was significantly
larger for the interacting than for the noninteracting galaxies and that
2) there was a shift in the distribution of the interacting galaxies towards
bluer B-V and U-B
relative to the noninteracting galaxies. Based on spectral evolutionary
models, their interpretation was that the major cause of both these effects
was a {\it significant increase in SFR} among the interacting
galaxies. In the most extreme cases the colours and luminosities correspond
to a consumption of most
of the available gas in less than a few times 10$^7$ yr. Following the work
by
LT,
a number of investigations have examined other criteria
of recent star formation and to a large extent confirmed the findings by LT
(e.g. Keel et al. \cite{keel}, Kennicutt et al. \cite{kennicutt2}, Bushouse
\cite{bushouse2},  Sekiguchi \& Wolstencroft \cite{sekig}).
These criteria include \ha, FIR and radio continuum but none of these
criteria is really univocally related to star formation so one should always
take care to discuss also other possible sources.

Below we will discuss the implications of the
broadband photometry, the spectroscopy of the central regions and the FIR
luminosities in the context of starburst activity.

   \begin{figure}
\sidecaption
\epsfig{file=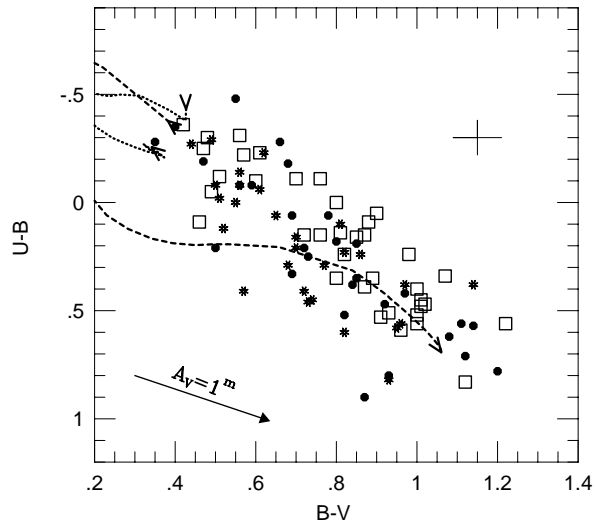, width=8cm}
%	\resizebox{\hsize}{!}{\includegraphics{H3242f6.ps}}
     \caption[]{The U-B/B-V diagram of the galaxies in the
present study based on the photoelectric photometry, corrected for galactic
extinction (Burstein \& Heiles \cite{burstein}). The filled symbols are
the components of interacting pairs, stars are merger candidates 
and squares are the
galaxies of the comparison sample. The apertures used in the photometry
correspond roughly to the effective diameter. For
comparison the evolutionary tracks of galaxies with two different star
formation histories are displayed, a 100 Myr burst (hatched line) and
continuous star formation (dotted line). We assumed
a Salpeter IMF and solar abundancies (from Zackrisson et al. 
\cite{erik}). The straight arrow is the reddening vector and the cross 
corresponds to 1$\sigma$ errors.}
         \label{ubbv}
   \end{figure}

\subsection{Optical data}
\subsubsection{Global properties}

\begin{table*}

\caption[ ]{Fluxes of the most prominent emission lines of the optical 
centres of the interacting and merging galaxies. The size of the aperture 
was 1.5"x3". The unit is 10$^{-19}$Wm$^{-2}$. Positive numbers are emission 
lines and negative are absorption lines.}
\begin{flushleft}
\begin{tabular} {lllllllll}
\noalign{\smallskip}
\hline
\noalign{\smallskip}
ESO-nr           & \ha  & \hb  & \hg & [OII]   &[O III]  & [N II]  & [S II]
&
[S II] \\
                 &  6563   &  4861   &  4340   &  3727   &  5007   &  6584

&
6717   &  6731  \\
\hline
                 &         &         &         &         &         &
&
&        \\
079-IG13 W     &   33 &    - &    - &    - &    - &    - &    - &    - \\
079- G16       & 2170 &  507 &  182 & 1180 & 1260 &  472 &  217 &  207 \\
080-IG02 W     &   31 &   11 &    4.0 &   28 &    6.8 &    5.6 &    4.7 &
4.0
\\
085-IG05       &   11 &    1.9 &    - &    6.1 &    3.3 &    2.9 &    3.7 &
   -
\\
105- G26 W     &    2.2 &   -0.6 &   -3.9 &    - &    - &    0.9 &    - &
1.2
\\
108-IG18 W     &  329 &   65 &   27 &    - &    - &  146 &   63 &   38 \\
108-IG18 E     &  156 &   27 &    - &    - &   50 &   46 &   48 &   27 \\
108-IG21       &   72 &   13 &    - &   25 &    5.1 &   24 &   26 &   - \\
110- G22 W     &   16 &    - &    - &    3.4 &    4.1 &    5.6 &  4.5 &  -
\\
110- G23 E     &   16 &    - &    - &    - &    - &   17 &    - &    - \\
117- G16       &   16 &    - &    - &    - &    - &    - &    - &    - \\
143- G04       &   86 &   16 &    - &    - &    - &   79 &    - &    - \\
145-IG21 N     &    3.1 &    - &   -1.5 &    - &    - &    3.3 &    - &    -
\\
145-IG21 S     &   44 &    6.7 &    2.3 &   11 &    4.5 &   20 &   14 &   -
\\
148-IG10       &   31 &    9.0 &    2.8 &   28 &    2.4 &   10 &   10 &   -
\\
151-IG36 W     &   37 &   11 &    - &   72 &   17 &    8.8 &   11 &   - \\
151-IG36 E     &   88 &    - &   -3.6 &    - &   74 &   90 &   33 &   - \\
157-IG05       &   18 &    4.3 &    - &    6.5 &    - &    4.5 &  4.3 &  -
\\
157-IG50 W     &   16 &    2.5 &    - &   16 &    8.0 &    1.0 &  3.4 &  2.1
\\
187-IG13 S     &    5.4 &    - &    - &    - &    - &    0.2 &    - &    -
\\
187-IG13 N     &    1.6 &    - &    - &    - &    - &    - &  1.4 &   - \\
188-IG18 W     &   25 &    1.6 &   -0.6 &    7.0 &    2.6 &   12 &    4.9 &
5.9 \\
188-IG18 E     &   70 &    8.6 &    2.9 &    9.0 &    5.4 &   29 & 30 &  -
\\
193- G19 N     &    4.0 &    - &    - &    7.3 &    - &   11 &   3.4 &  2.2
\\
199- G01       &  105 &    - &    - &    - &    - &   55 &   12 &   16 \\
200-IG31 N     &   40 &    - &    - &    - &    - &    - &    - &    - \\
200-IG31 S     &    6.6 &    - &    - &    - &    - &    - &    - &    - \\
205- G01       &  101 &   30 &   16 &  102 &  105 &    3.3 &  7.7 &  5.3 \\
235-IG23 N     &    5.2 &    - &    - &    - &    - &    - &    - &    - \\
243- G15 N     &    - &   -5.6 &   -2.7 &    - &    - &    3.8 &  11 &  - \\
244- G12 N     &  390 &   29 &    5.9 &   21 &   26 &  214 &   48 &   40 \\
244- G12 S     &   26 &    7.4 &    - &   16 &    5.6 &   12 &  5.1 &  4.5
\\
244- G17 W     &   -0.2 &    1.9 &   -0.6 &    - &    5.7 &    2.4 &    - &
   -
\\
244- G17 E     &  133 &   32 &   13 &    7.3 &   36 &  -11 &   4.4 &  3.5 \\
244-IG30       &   33 &    7.5 &    - &   37 &   11 &   10 &  11 &  9.0 \\
244- G46 E     &   -1.5 &   -6.8 &    - &    - &    - &    2.7 &    - &    -
\\
249-IG31       &  116 &   28 &    9.1 &   75 &   60 &   18 &    15 &  5.5 \\
284-IG41 S     &   11 &   -6.7 &   -1.0 &    - &    - &    6.8 &  5.7 &  -
\\
284-IG48       &    6.8 &    - &   -1.4 &    - &    - &    2.1 &    1.6 &
0.6
\\
285-IG35       &    3.2 &   -0.8 &    - &    8.5 &    - &    2.3 &    - &
-
\\
286-IG19       &  245 &   41 &    - &  154 &   46 &   89 &   75 &   53 \\
287- G40       &    - &   -0.5 &    - &    - &    - &    1.3 &    - &    -
\\
288- G32 W     &   22 &    3.2 &   -0.7 &    7.7 &    2.7 &    6.7 &  6.8 &
-
\\
290- G45       &   76 &   11 &    - &   24 &    9.2 &   26 &   27 &   - \\
293- G22 N     &   47 &    - &    - &    - &    - &   26 &    - &    - \\
297- G11 W     &   97 &   10 &    - &   12 &    6.6 &   43 &    12 &   8.9
\\
297- G12 E     &   50 &    5.3 &    - &    8.8 &    5.3 &   28 &  11 &   27
\\
299-IG01 S     &   25 &    3.2 &    1.5 &   12 &    2.7 &   16 &   4.8 &
1.7
\\
306- G12 S     &    4.8 &    - &   -1.2 &    - &    - &    1.6 &    - &    -
\\
340- G29       &   19 &   -3.8 &    - &    - &    - &    6.6 &    6.5 &
2.7 \\
342-IG13 N     &   66 &    8.6 &   -6.8 &   39 &   14 &   20 &    13 &   9.6
\\
344- G13 S     &   10 &    1.4 &    - &    - &    - &    2.7 &   - &  5.1 \\

285-IG35       &    3.2 &   -0.8 &    - &    8.5 &    - &    2.3 &   4.7 &
1.1
\\
\noalign{\smallskip}
\hline
\end{tabular} \\
\end{flushleft}
\end{table*}

\begin{table*}

\caption[ ]{Fluxes of the most prominent emission lines of the
galaxies in the comparison sample. The unit is 10$^{-19}$Wm$^{-2}$.
Positive numbers are emission lines and negative are absorption lines. Only
galaxies with
detected emission lines are included.}
\begin{flushleft}
\begin{tabular} {lllllllll}
\noalign{\smallskip}
\hline
\noalign{\smallskip}
ESO-nr           & \ha  & \hb  & \hg & [OII]   &[O III]  & [N II]  & [S II]
&
[S II] \\
                 &  6563   &  4861   &  4340   &  3727   &  5007   &  6584
&
6717   &  6731  \\
\hline
                 &         &         &         &         &         &
&
&        \\
015- G05       &    9.9 &   -5.1 &   -2.3 &    - &    - &   23 &    - &    -
\\
026- G04       &    - &   -1.1 &    - &    - &    - &   11 &    - &    - \\
027- G14       &   35 &    5.4 &    - &    - &    - &   14 &    3.9 &    1.9
\\
048- G25       &    7.9 &    - &    - &    - &    - &    1.2 &    - &    -
\\
074- G26       &   17 &    - &    - &    - &    - &    6.8 &    - &    - \\
106- G08       &   10 &    - &    - &    - &    - &    5.5 &    - &    - \\
151- G43       &    2.8 &    - &    - &    - &    - &    0.3 &    - &    -
\\
153- G33       &    3.5 &    - &    - &    - &    - &    2.9 &    1.1 &
0.8
\\
201- G12       &   13 &    - &    - &    - &    - &    - &    - &    - \\
200- G36       &   22 &   -4.6 &    - &    - &    - &   12 &    4.3 &    2.9
\\
240- G12       &   21 &    3.6 &    - &   14 &    5.9 &    7.0 &    7.6 &
2.7
\\
287- G21       &    8.7 &    - &    - &    - &    - &    - &    - &    - \\
293- G04       &   11 &    - &   -8.1 &   24 &    3.1 &    3.3 &    2.5 &
1.9
\\
299- G07       &  269 &   30 &    - &    - &    - &  128 &   24 &   21 \\
299- G20       &  269 &   33 &   13 &    - &    5.0 &  139 &   22 &   20 \\
303- G14       &   79 &   13 &    - &   11 &    7.6 &   48 &    11 &   7.6
\\
340- G07       &    6.0 &    - &    - &    - &    6.9 &    9.0 &    - &    -
\\
341- G32       &    4.8 &    - &    - &    - &    - &    - &    - &    - \\
345- G49       &   11 &   -2.0 &    - &    - &    - &    4.4 &    - &    -
\\
\noalign{\smallskip}

\hline
\end{tabular} \\
\end{flushleft}
\end{table*}

   \begin{figure}
\centering
%\includegraphics[width=12cm]{ubbvhist.ps}
%      \vspace{0cm}
%	\resizebox{\hsize}{!}{\includegraphics{H3242f7.ps}}
\epsfig{file=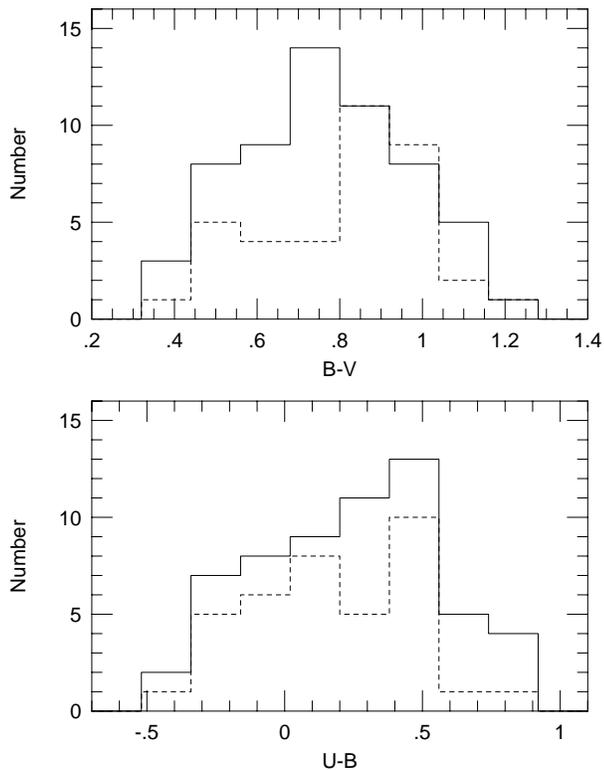, width=8cm}
      \caption[]{The cumulative distributions of U-B and B-V. The
full-drawn line are the interacting and merging galaxies and the
hatched line is the comparison sample.}
         \label{ubbvhist}
   \end{figure}

Fig. \ref{ubbv} shows the two-colour diagram based on our photoelectrically
obtained UBV data and Fig. \ref{ubbvhist} shows the corresponding cumulative
distributions. The colours have been corrected for Galactic extinction
according to Burstein \& Heiles (\cite{burstein}). The
apertures used in the photometry have been chosen so that they approximately
correspond to the effective diameters. The error cross in the figure is the
estimated total standard deviation in the colours. The estimate is based 
on
a few unrelated factors. We estimate the internal uncertainty due to
instrumental effects to about 0.02 mag. To this is added the effect due 
to 
the problem with
the photometric quality of the night, the accuracies of the stardard stars, 
the centering of the aperture and the fact that we do not make any effort to
homogenize our data to a standard diameter. The effects of the first
three problems can be estimated from the data we have obtained here and from
other investigations we have carried out, where we have collected many 
observations
of the same galaxy in the same diaphragm. An independent check of the
consistency of our data is obtained from comparisons with results from
other groups. Table 1 lists all galaxies in our sample for which there
exists comparable photometry from other studies. In general the agreement is
very good. One strongly deviating case is ESO 236-IG01. We cannot judge what 
could be the explanation of this and we therefore exclude it in the discussion.
The median difference between our remaining data and data from other
groups are 0.01 mag. in U-B and 0.02 in B-V and the median of the individual 
deviations are 0.04 mag. in both colours. The morphological type dependent
relations between colours and aperture/effective diameter, presented
in diagrams in RC3 (de Vaucouleurs et al.
\cite{dev4}) was used to obtain estimates of the uncertainty in the
colours due to the fact we do not use a standard diameter. We finally
adopted $\sigma$(U-B)=0.08 and $\sigma$(B-V)=0.06.

   \begin{figure}
\sidecaption
%\includegraphics[width=17cm]{arp1.ps}
%       \vspace{0cm}
\epsfig{file=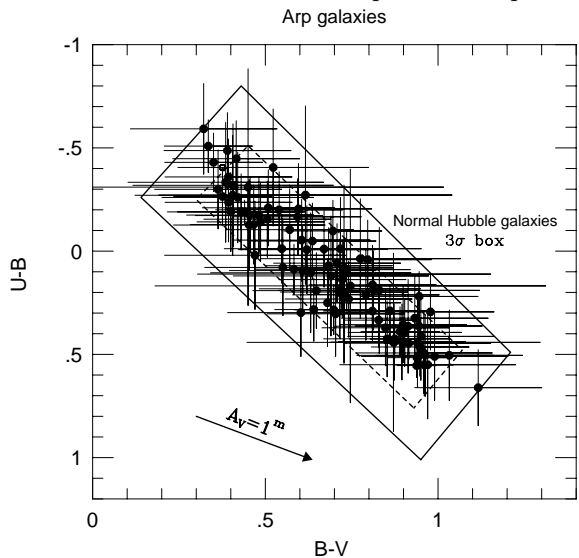, width=8cm}
%	\resizebox{\hsize}{!}{\includegraphics{H3242f8.ps}}
      \caption[]{Arp galaxies with available photometry in NED (filled dots), 
      corrected for galactic extinction (Burstein \& Heiles \cite{burstein}). 
      The error bars are observational mean errors obtained from NED. The small 
hatched box 
      indicates the location of
normal Hubble galaxies (de Vaucouleurs \cite{dev3}). The size of the box 
corresponds to 3 standard deviations from the mean colours. The larger solid box 
is the Hubble box with a size corresponding to 3 times the median value of the 
observational mean errors of the Arp sample.
The long arrow is the reddening vector.}
         \label{arp1}
   \end{figure}

   \begin{figure}
%\sidecaption
%\includegraphics[width=17cm]{H3242f9.ps}
%      \vspace{0cm}
\epsfig{file=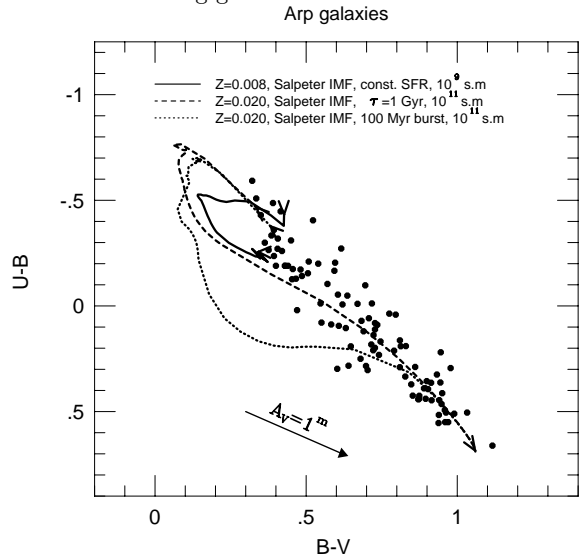, width=8cm}
%	\resizebox{\hsize}{!}{\includegraphics{H3242f9.ps}}
      \caption[]{Arp galaxies also shown in Fig. 6 (filled 
dots),
corrected for galactic extinction (Burstein \& Heiles \cite{burstein}). For
comparison the evolutionary tracks of galaxies with three different star
formation histories are displayed, an exponentially decaying star formation 
rate
with a timescale of 1 Gyr (full drawn line) a 100 Myr burst followed by
passive evolution (hatched line) and
continuous star formation (dotted line). The start and end of the tracks 
are marked with arrows. We assumed
a Salpeter IMF and solar abundancies (from Zackrisson et al. 
\cite{erik}). The straight arrow is the reddening vector.}
         \label{arp2}
   \end{figure}
   
For comparison Fig. \ref{ubbv} also
displays the predicted evolution of a dust free star forming galaxy with
solar abundances and a Salpeter initial mass function (Zackrisson et al.
\cite{erik}). Two extreme star formation scenarios were assumed: 1) a burst 
with a duration of 10$^8$ yr and 2) a continuous star formation. As is seen, 
the predicted
colours agree reasonably well with the mean colours of the samples. As
expected, the model with a short burst followed by passive evolution fits the 
galaxies of elliptical type while the model with
continuous star formation fits better the irregulars and spiral galaxies.
But no model within a metallicity range of 0.01-2 times solar can explain the
colours in the lower (large values of U-B) left part of the distribution. A
comparison with predictions from other models (Fioc \& Rocca-Volmerange
\cite{fioc1}, \cite{fioc2}; Worthey \cite{worthey}) can not remedy the
situation. We note that similar, "deviating" colours are also found in 
e.g. the sample of spiral galaxies by Gavazzi et al. (\cite{gavazzi}).

{\it Somewhat surprisingly, we  do not see the same trends as seen in the LT
data}. As in their corresponding diagram, the IGs also here have a larger
dispersion than the NIGs but only slightly so. What is more
interesting is that instead of a blue excess, the envelope of the IGs shows
a
small red excess basically in U-B relative to the noninteracting galaxies.
Why is our result different from that of LT? 

There may be a suspicion that a difference in mean absolute luminosity 
between our two samples would introduce a systematic difference in the colours 
according to the well established colour-luminosity relationship.
Later we will indeed claim that the LF of the IGs extends to significantly 
higher 
luminosities than the NIGs. Luminous galaxies tend to be redder than less 
luminous ones so if we compare galaxies drawn from a normal sample, the mean 
colours would correlate with mean luminosities. Whatever the reason may be, 
it is interesting to ask whether it would be possible that this effect could 
hide a blue excess caused by increased star formation triggered by the 
interaction. 

We have estimated the effect of this possible bias from the study of normal 
galaxies carried out by Jansen et al. (\cite{jansen}). From Fig. 5c in their 
paper we estimate the trend in B-R to d(B-R)/dM$_B\sim$0.07, for both 
early and late type
galaxies. The mean magnitude difference between the NIG sample and the IG 
sample is $\Delta M<$ 1 mag, which would result in a difference of 
$<$0.07 mag. in B-R. Assuming the corresponding difference in B-V to be 
approximately the same and in U-B to be slightly smaller (Jansen's et al. 
Fig. 5d), a possible effect of differences in absolute magnitudes would shift 
the distributions of the IGs and the NIGs in the two-colour diagram to make 
them come somewhat closer but not to change the relative positions. This is 
completely insignificant for the results. In addition we would like to point 
out that our results in this paper indicate that interacting galaxies are 
likely to have experienced more mergers than isolated galaxies. This would 
tend to reduce differences in colours betweeen galaxies of different 
luminosities when compared to the monolithic evolutionary scenario thus making 
the estimated colour difference even smaller.

Let us for a while assume that the IGs have a significantly increased star 
formation. Could the effect be masked by reddening effects due to different 
distribution and amount of dust in the NIGs vs. the IGs? It cannot exclusively 
be
due to an increased effect
of dust relative to the NIG sample since the slope of the envelope for the
internal reddening is steeper than the reddening vector in the normally
assumed "screen" dust distribution. Also, the effect of dust would be to
shift the sample of interacting galaxies above, not below that of
the noninteracting galaxies in the two color diagram.
The effects of dust on broad-band colors may even be smaller than often
suggested, because of the wavelength-dependent effects of dust absorption
(reddening) and scattering (blueing), which may largely compensate each
other
in a dusty interstellar medium (Mathis \cite{mathis}). A detailed study of
the
effects of dust on broad-band colors has been carried out by Witt et al.
(\cite{witt}), who estimated the reddening effects by including scattering
of
light and making the calculations for star and dust distributions imitating
different astrophysical environments. Witt et al. point out the fact that a
small reddening will be obtained both in galaxies with a small amount of
dust
{\it  and} in very dust rich starburst galaxies where the dust is centrally
concentrated such that it hides most of the light from the central
starburst.
The latter situation may occur if the dust during a close interaction or in
a
merger is redistributed such that a substantial part of it is brought to the
centre.  This would clean up the "off-centre" regions and make the centre
more
opaque. This is a possible explanation of the different distributions of the
IGs and NIGs in our two-colour diagram.

But let us return to the comparison with the LT results. There are
important differences between their sample and ours. Firstly, the galaxies
in
our sample were obtained from a catalogue limited to a region at Galactic
latitude $\leq$ -30$^\circ$, with the purpose to avoid problems of Galactic
extinction. The lower limit in Galactic latitude by LT was 20$^\circ$ and
thus
Galactic extinction starts to become a problem. However, LT did not correct
for  Galactic extinction. According to them Galactic extinction could not
cause the dispersion in the diagram, because the
reddening line is in the same direction as the slope of the distribution of
the galaxies in the diagram (internal reddening is supposed to be even less
important but this may be debated). However, the lack of attention to the
effects of Galactic extinction at low latitudes certainly can explain part
of the differences between LT and us.

A second reason for the disagreement with LT and us may be the
differences in sample homogeneity. Our samples were selected from a
magnitude
limited catalogue compiled by the same person and the photometry was
obtained
by the same team using the same instrumentation. On the other hand, while
the Hubble galaxies are all nearby well studied galaxies, often having
detailed
photometry out to large distances, the Arp objects are significantly more
distant and have less frequently good photometric data available. One would
suspect that this could induce an increased scatter
in the Arp sample. LT claim that the mean error in the measurements is more
or
less the same in the two samples. However, a remarkable fact is that
the photometric errors of the individual Hubble galaxies are significantly
larger (about a factor of 3) than the scatter in the two-colour diagrams
by LT.

An additional contribution to the differences found by LT and not by us may
lie in the selection of galaxies. The LT sample is biased towards the most
dramatic cases of closely interacting galaxies and mergers and thus is not
representative for this type of galaxies. They also include
blue compact galaxies, Seyfert galaxies and "peculiar" galaxies in
general.
They do not discuss the
effects of differences in the distribution of morphological types in the two
samples but from their diagrams one gets the impression that there is an
underrepresentation of late type galaxies in the Hubble sample. This may
have
an important impact on their result.

The results found by LT were also discussed by Bergvall (\cite{bergvall3})
who compared the colours of noninteracting galaxies taken from RC2, based on
statistics of a larger sample than the Hubble galaxies, with a sample of
galaxy pairs, different from our sample here. He found that the difference
between interacting (based on data from RC2) and noninteracting galaxies as
regards the dispersion perpendicular to the correlation in the colour-colour
diagram disappears. Bergvall's
conclusion was that there was no difference between interacting and
noninteracting galaxies as regards the scatter in the U-B/B-V diagram.

We have reconstructed the diagram of UBV colours of Arp galaxies as based
on the total magnitudes available in NED. We show the result in Fig.~\ref{arp1}.
All data have been corrected for galactic extinction according to Burstein
\& Heiles (\cite{burstein}).
The NED photometry of NGC 1241 disagrees strongly
with more recent data from LEDA and this galaxy was not included in the 
diagram.
NGC 1596 was also not included because of the large uncertainty in the
extinction correction. We note differences between our diagram and Fig. 1. 
of the work by LT. The spread of the data is smaller in our diagram and the 
distribution is more gaussian with the data more concentrated towards the mean 
Hubble distribution, especially at the blue and red edges. This fact is a
consequence of an increased quality in the data and supports our suspicion that 
the
error bars in the LT diagram are too small, leading to the erroneous
conclusion that the intrinsic scatter of the Arp galaxies is dramatically
larger than that of the normal Hubble galaxies. The error box in the LT Fig. 1 
corresponds to a mean error in both colours of $\sim$ 0.04 magnitudes. As they 
say in the figure caption, this is obtained from RC2. This is hard to understand 
since the median error of the corresponding sample, of similar size, obtained 
from RC3 is $\sigma \sim$ 0.13 magnitudes. Once we adapt the RC3 value, that 
should be more correct, the data points of all galaxies in
our Fig.~\ref{arp1} are within the quoted errors. No extra scatter is needed to 
explain the distribution perpendicular to the major trend. 

To clarify this point we have indicated with a box (the small hatched box) 
in the diagram the location of normal Hubble galaxies 
according to de Vaucouleurs (\cite{dev3}). The galaxies are obtained from RC2 
and the colours have been corrected for galactic extinction. Originally the 
colours were also corrected for the redshift (the K-effect) but we have 
readjusted these corrections to correspond to the median redshift of the Arp 
galaxies (4950 kms$^{-1}$) according to the same procedure as used by de 
Vaucouleurs in RC2 (\cite{dev}). Before including the data to calculate mean 
colours, de Vaucouleurs rejected those that deviated more than 2$\sigma$, thus 
restricting the final data sample to galaxies showing normal behaviour. The 
width of the box corresponds to $\pm$ 3 standard deviations of the mean colors 
(3 x de Vaucouleurs' $\sigma(\xi_2)$ parameter). To compare with the Arp 
galaxies we also display in the diagram the same 'Hubble box' but increased to a 
size that corresponds to the median observational errors of the Arp galaxies 
($\sigma$(U-B)=$\sigma$(B-V)=0.13 mag.). If
there is no extra scatter among the Arp galaxies compared to normal Hubble 
galaxies and the distribution is
gaussian we thus expect almost 100\% of the galaxies to be located within the 
box. This is in fact the case. Thus there is no support from the global 
broadband colours of an increased star formation activity triggered by the 
interactions. We also note that the distribution of the
Arp galaxies agree well with our data but that the scatter is slightly
larger for our galaxies. This may be a consequence of a larger morphological
diversity among our data and the fact that our data have not been 
homogenized to total magnitudes whereas the NED data are homogenized.

Fig.~\ref{arp2} shows the Arp data in comparison with evolutionary tracks
derived from our spectral evolutionary models. We find that the observations
agree very well with the model predictions. As shown also in fig.~\ref{ubbv},
we find no galaxies along the
first part of the evolutionary tracks. This is probably a consequence of
internal extinction. In our studies of luminous blue compact galaxies we
have found that the extinction we derive from the H$\alpha$/H$\beta$
emission line ratio typically amounts to a little less than A$_V$=1
magnitudes. If we apply this amount of extinction to the late type galaxies
in the diagram the evolutionary track will follow the right hand side of the
envelope of the colour distribution of the Arp sample. We conclude that all
galaxies in the diagram would fit into the traditional model of galaxy
evolution where the stellar population has a Salpeter IMF and where massive
galaxies consume the gas faster than the low mass galaxies. The scatter of
the UBV data gives no support of interaction induced star formation and
definitely not of starbursts.

The reason we have devoted so much space for discussion of the now rather
old result by LT is that it still has such an impact on our concept of
tidally induced star formation.  We conclude that {\it there is no strong
support of a significant difference between the global optical colours (and thus
neither in global SFRs) of noninteracting and
interacting/merging galaxies} in our samples nor in the Arp sample with 
available photometry.

   \begin{figure}
\epsfig{file=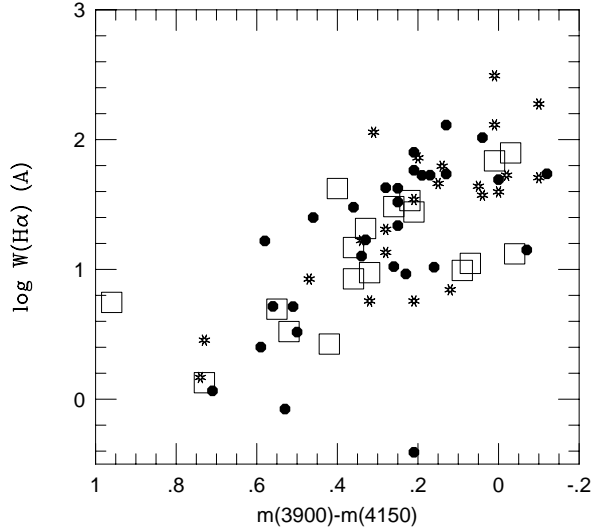, width=8cm}
%	\resizebox{\hsize}{12cm}{\includegraphics{H3242f10.ps}}
      \caption[]{The m$_{3900}$-m$_{4150}$ colour plotted against
the equivalent width of \ha. The symbols are the same as in Fig.
\ref{ubbv}.}
         \label{ubx}
   \end{figure}

\subsubsection{Nuclear properties}

Even if we see no tendency for increased star formation in the global
colours, the colours of the
very central regions behave differently. To obtain the colours, we 
integrated the spectra using
a top-hat transmission function with a width of 100 \AA. We study two
colours.
The first is m$_{4150}$-m$_{5500}$, roughly corresponding to B-V. The second
is based on the Balmer jump, represented by m$_{3900}$-m$_{4150}$. The
advantage with using this colour is that it is only weakly dependent on
extinction (typically $<$ 0.1$^m$). The m$_{3900}$-m$_{4150}$ index grows 
monotonically with
spectral
type and is a measure of star formation activity. This can be seen in Fig.
\ref{ubx}, showing the colour plotted against the equivalent width of \ha.
We
see a clear correlation between the two parameters. We also see that \ha
emission is mostly found in objects with blue colours, m$_{3900}$-m$_{4150}$
$\le$ 0.4.

   \begin{figure}
\epsfig{file=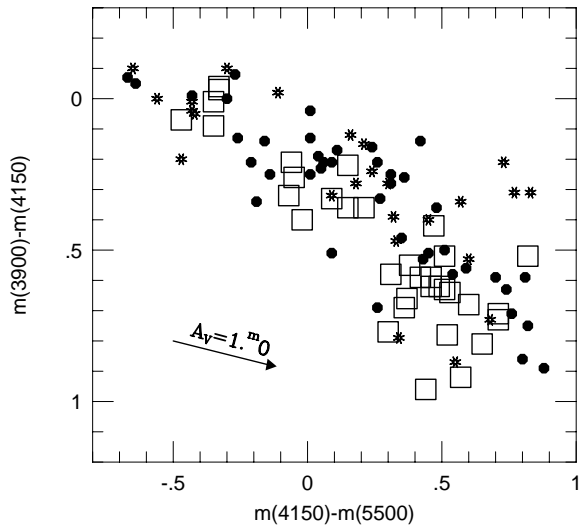, width=8cm}
%	\resizebox{\hsize}{12cm}{\includegraphics{H3242f11.ps}}
      \caption[]{The m$_{3900}$-m$_{4150}$/m$_{4150}$-m$_{5500}$ two-colour
diagram. The symbols are the same as in Fig. \ref{ubbv}.}
         \label{ubbvx}
   \end{figure}

Figure \ref{ubbvx} shows the two-colour diagram of the galaxies. Here one
clearly sees that, as m$_{4150}$-m$_{5500}$ becomes redder, a significant
difference in the distribution appears between IG and NIG. A possible
explanation is that the IGs have experienced a major star formation event
more
recently than the NIGs which would shift the 3900-4150 index towards the
blue. Another explanation could be that the young stars in the IGs are more
centrally concentrated. A third possibility is that the colours of the IGs
are
more affected by dust as the increased star formation activity is located
mainly in the central regions. Probably all three effects contribute.

\begin{figure}

\epsfig{file=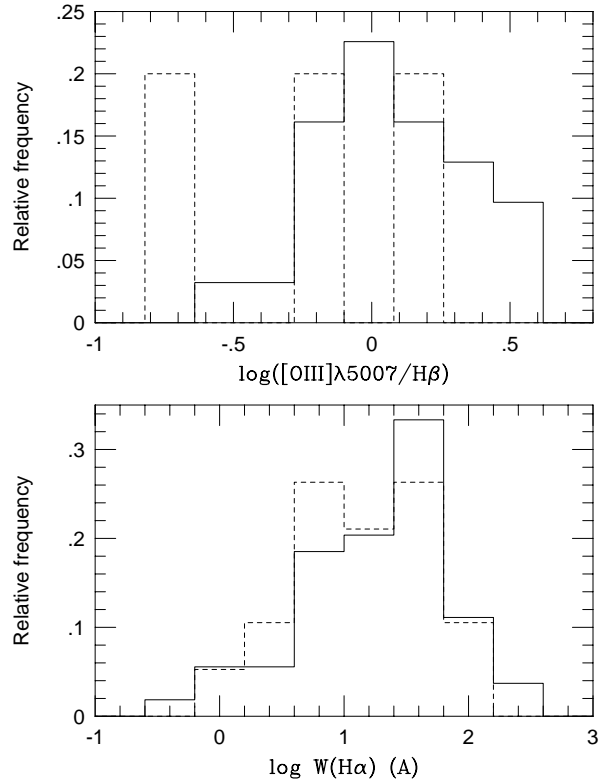, width=8cm}
%	\resizebox{\hsize}{12cm}{\includegraphics{H3242f12.ps}}
      \caption[]{The frequency distribution of the
[OIII]$\lambda$5007/\hb line ratio
and
the equivalent width in \ha. The solid line is the
interacting
and merging galaxies and the hatched line is the comparison sample (only 3
galaxies in the upper diagram).
Corrections for Galactic extinction have been applied.}
         \label{o3}
   \end{figure}

It is interesting to compare this result with two spectral line parameters,
\ha and [OIII]$\lambda$5007/\hb, shown in Fig. \ref{o3}. These data were
derived
from the same spectra of the central regions as the data in Fig.
\ref{ubbvx}.
In Fig. \ref{o3} the same trend as in the previous diagram is seen - a
support
for an increased SFR in IGs. The relative number of galaxies with measurable
[OIII] lines are much higher in the IG sample than in the NIG sample (34 and
13\% respectively). 58\% of the IGs and 47\% of the NIGs show \ha. A
Kolmogorov-Smirnov
test of the equivalent width of \ha (\wha) shows that the probability that
the
two samples are drawn from the same population is 17\%. Thus, even if the
support is
weak, central star formation is more frequent in interacting and merging
galaxies, confirming the previous findings. Our result is also in the same
line
with the previous study by Laurikainen \& Moles (\cite{laurik3}),  who used
\ha intensity to show that star formation in interacting systems in
comparison
to isolated galaxies is pronounced mainly only in the central regions of
the galaxies. The increase in SFR we find in our sample is modest however.
The mean \wha is 40 \AA ~and 16 \AA ~for the IGs and NIGs, respectively. 
This result agrees well with two other investigations of the nuclear (i.e. 
non-global) properties. One is the study of the difference between Arp galaxies 
and isolated spiral galaxies by Keel et al. (\cite{keel}) and the other a study 
of interacting pairs by Donzelli \& Pastoriza (\cite{donzelli}). None of these 
investigations were however based on magnitude limited samples. 

The difference in \wha between IGs and NIGs that we obtain
corresponds to a difference in SFR of a factor of 2-3 in the
mean. Thus it is not relevant to talk about starbursts except for a few
cases. Possible effects due to differences in luminosity functions could 
change these numbers with not more than about 10\% (Carter et al. 
\cite{carter}). 
Although a direct comparison is not possible due to the difference in the
contribution from old stars to the stellar continuum, true starbursts as 
those one may witness in some blue compact galaxies (cmp. Terlevich et al. 
\cite{terlevich}), have \wha much larger, of the order of several 100 \AA.  
Moreover, the [OIII]$\lambda$5007/\hb
ratios of the dwarf starbursts are several times higher than in our sample
of IGs. We may compare the increase in SFR obtained from \wha with that 
derived from radio continuum data. Hummel et al. (\cite{hummel2}) studied a 
large sample of spiral galaxies and found that those involved in interaction 
had a nuclear radio continuum emission 5 times stronger than the more 
isolated ones and that the radio emission was correlated with the activity 
in HII regions. Taking into account that the IGs appear to have a higher gas 
mass (see the discussion below), and assuming the star formation rate being 
correlated with the gas mass, there seem to be an agreement between our result 
and that of Hummel et al..

Another relevant issue is the frequency of AGNs. The increased
level of excitation in the interacting galaxies seems not to have resulted
in
Seyfert-type activity in them. By using the criteria by Osterbrock
(\cite{osterbrock}) we found only 2 Seyferts (244-G17W and 151-IG36E) among
the interacting galaxies, which means 4\% , and none in the comparison
sample.
The frequency of 4\% we obtained is similar as found for samples of 
interacting and noninteracting galaxies in previous investigations. Dahari, 
\cite{dahari2}, 
and Keel et al. \cite{keel} used a field galaxy sample composed of the samples 
by Keel (\cite{keel2}) and Stauffer
(\cite{stauffer}) and found that the contribution of Seyferts
in the NIG-sample was 4.6\% and 5.6\% respectively, the fraction of Seyferts in 
IG samples being rather similar. Barton et al. (\cite{barton}),
using a different sample, found that 4\% were Seyferts. On the other hand, our 
comparison sample is so small that in the same frequency level of active 
galaxies no Seyferts would be expected. 16\%
of the IGs display LINER like properties with [NII]$\lambda$6584/\ha $\geq$
0.7  and [OIII]$\lambda$5007/\hb $\leq$ 3. Within the statistical
uncertainties,
this is equal to the relative number of LINERs in an apparent magnitude
limited
general sample of galaxies ($\sim$20\%; Ho et al. \cite{ho}). There are too
few
NIG spectra available to allow for a statistically reliable estimate of this
number.

\subsection{Near-IR data}

\begin{figure}
%\sidecaption
%\includegraphics[width=17cm]{H3242f13.ps}
\epsfig{file=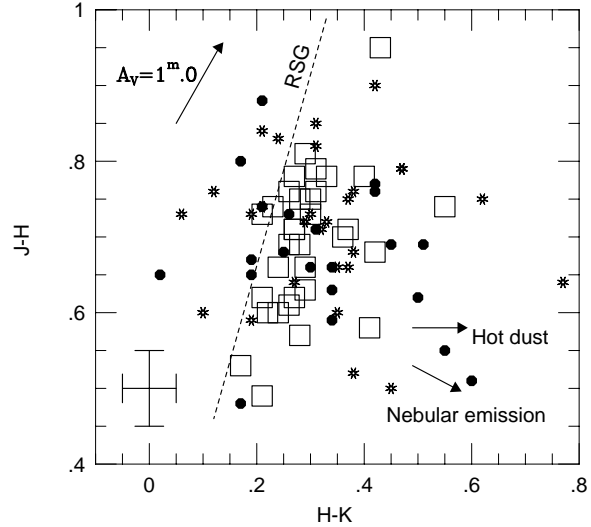, width=8cm}
%	\resizebox{\hsize}{!}{\includegraphics{H3242f13.ps}}
      \caption[]{The J-H/H-K two-colour diagram. The symbols
are the same as in Fig. \ref{ubbv}. The position of red
supergiant stars has been indicated with the hatched line marked RSG.
The cross at the lower left indicates 1$\sigma$ errors.}
\label{jhhk}    
\end{figure}

Fig. \ref{jhhk} shows the near-IR J-H/H-K two-colour diagram of the galaxies
that were sufficiently bright to measure. The observations were carried out
with apertures that cover a substantial part of the galaxy and are in this
sense comparable to the optical data for global galaxy properties. But we
will see that a larger portion of the near-IR emission probably originates
from the nuclear region than the optical data do. Fig. \ref{jhhk} shows that
while the comparison galaxies are contained within a well defined sequence
close to the distribution of red supergiants, the interacting galaxies show
a
significantly larger scatter. Since this cannot be due to dust extinction
(cf.
the reddening arrow in the diagram), the reason has to be found in the
properties and spatial distribution of the stellar population and the
emission
from circumstellar and interstellar gas and dust and possible AGNs. AGNs are
normally situated in the right hand part of the diagram (e.g. Spinoglio et al.
\cite{spinoglio}). The mean H-K of Seyfert 1:s is $\sim$ 0.84$\pm$0.33, i.e.
significantly separated from the envelope of the IGs. The mean H-K of
Seyfert
2:s
is $\sim$ 0.61$\pm$0.37 which is just at the edge of the IG distribution.
The effects of emission from hot dust and nebular emission
have been indicated in the figure.  It seems likely that the spread of the
IGs can be traced to such processes, possibly mixed with weak AGN activity.

The fact that the photometric properties of the IG and NIG samples differ
strongly in the near-IR but not in the optical seems to indicate that a 
substantial part of the energy production in IGs is taking place in regions 
that are heavily enshrouded by dust. This is not necessarily an indication 
of hidden
starburst activity. One should also consider effects of shock heating caused by
infalling clouds, possible hidden AGN activity and the consequences of
different distribution of gas and dust relative to the stars and other heat
sources in the two samples.

\subsection{Far-IR data}

\begin{figure}
  % \sidecaption
%\includegraphics[width=17cm]{H3242f14.ps}
	\epsfig{file=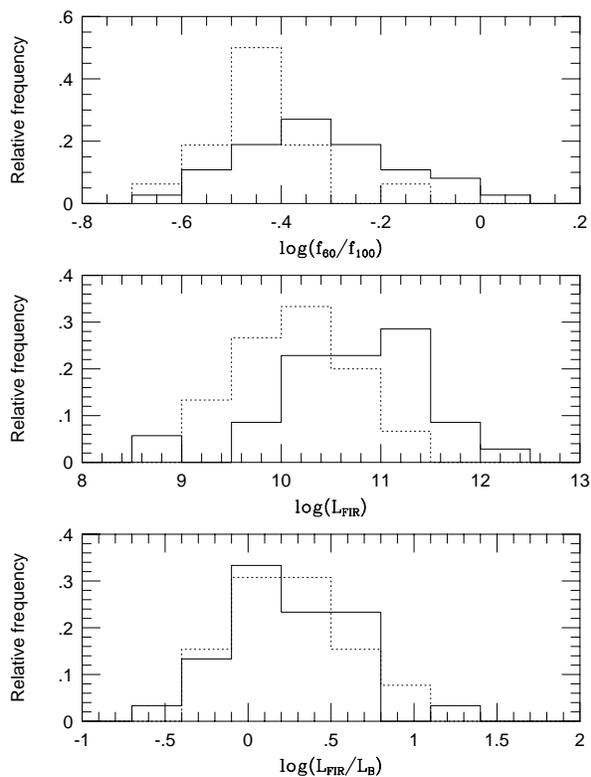, width=8cm}
%\resizebox{\hsize}{!}{\includegraphics{H3242f14.ps}}
      \caption[]{a) The distribution of the IRAS temperatures
based on the 60 and 100 $\mu$ fluxes. b) The IRAS luminosities in solar 
units. c) The ratio between the FIR luminosity and the blue luminosity. The 
solid line is the interacting and merging galaxies and the dotted line is 
the comparison
sample.}
         \label{iras}
   \end{figure}

Figs. \ref{iras}a and \ref{iras}b show the histograms
of the FIR luminosities and the dust temperatures
of the galaxies in the IG and NIG samples. The luminosities and temperatures
are based on the IRAS 60 and 100 $\mu m$ fluxes from the Faint Source Catalogue 
(Moshir et al. \cite{moshir}). When these fluxes were used in 
combination with 
optical/near-IR data in the discussion below, they were corrected for IRAS 
aperture confusion - thus the total magnitudes of interacting pairs were 
used when the IRAS aperture covered both galaxies of the pair. Absolute 
luminosities were calculated from the equation
L$_{FIR}$ = 5.5e5 r$^2$ (2.58f$_{60}$+f$_{100}$) L$_{\odot}$ (Lonsdale \& 
Helou \cite{lonsdale}) where r is the distance in Mpc and f is in Jy:s. We 
see that there is a significant difference both in the distribution of the 
IRAS luminosities and in mean dust
temperatures between the samples. The IGs are brighter and
hotter than the NIGs thus confirming the results from previous studies of
interacting galaxies (Young et al. \cite{young1}, Kennicutt \& Keel
\cite{kennicutt1},
Bushouse et al. \cite{bushouse1}, Telesco et al. \cite{telesco},
Laurikainen \& Moles \cite{laurik3}, Xu \& Sulentic \cite{xu}, Liu \&
Kennicutt \cite{liu}). One would normally
draw the conclusion that this is
due to an increased SFR and/or {\it efficiency} in interacting
galaxies. However,
Fig. \ref{iras}c hints at an alternative explanation. The diagram shows the
ratio between
the FIR and blue luminosities (based on aperture photometry approximately
at the effective radius). Contrary to what one would expect, judging
from results from previous studies (e.g. Liu \& Kennicutt \cite{liu}), the
distribution
is essentially the same for the two samples. We applied a Kolmogorov-Smirnov
test to the data in the three cases shown in the diagram and found that the
probabilities that the samples are drawn from the same population are 0.6\%
when the IRAS temperatures were tested, 0.3\% for the luminosities and 34\%
for
the FIR/B data. Thus, while the distributions of the 60/100 ratios and the
IRAS luminosities are significantly different, {\it the FIR/B ratios are
not}.

In the previous section we concluded that the star formation activity
is significantly higher only in the central region of IGs as compared to
NIGs. However, even if one assumes that the difference in the nuclear \wha 
between IGs and NIGs represents a difference in global SRF, the measured 
increase is insufficient to account for the enhanced IRAS 60 and 100 $\mu$m 
fluxes in the IG sample. The difference in dust temperature can explain part 
of the increase in luminosity but rather modest. If we assume an emissivity 
of $\epsilon \propto$ B$_\nu \lambda^{-\beta}$, with 1.4$\le \beta \le$2.0 
(Lisenfeld et al. \cite{lisenfeld}), only a 60\% increase of the enhanced 
FIR emission can be explained. Thus, the difference between IGs and NIGs in 
the FIR flux may not primarily be due to different SFRs but to different 
mass and different distribution of stars and dust. On top of this there may 
be additional
energy sources. These include hidden AGNs and mechanical heating when the 
potential energy is released from infalling gas clouds. Those sources are 
expected to be important in massive systems but are likely to play a minor 
role here. It appears that the main difference between the NIGs and the IGs 
is the scale in mass. One possible explanation is that, although originally 
having a
similar luminosity function as the NIGs, IGs now have become  brighter than
NIGs because they are located in a richer environment so that they merge
more frequently and thus build up the masses faster.

\begin{figure}
      \vspace{0cm}
\epsfig{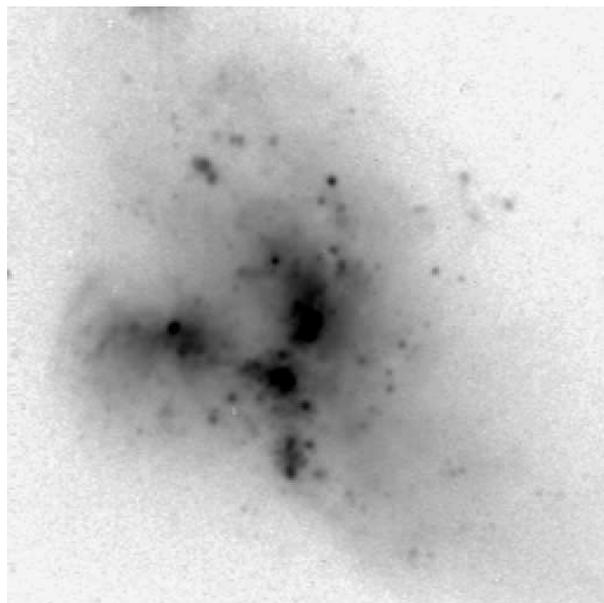}
%	\resizebox{\hsize}{8cm}{\includegraphics{H3242f15.ps}}
      \caption[]{NGC 1487 in B filter. The size of the field is 2.2'x2.2'.  
      North is up, east is to the left. ESO/MPI 2.2-m telescope, La Silla.}
         \label{N1487B}
   \end{figure}

\begin{figure}
      \vspace{0cm}
\epsfig{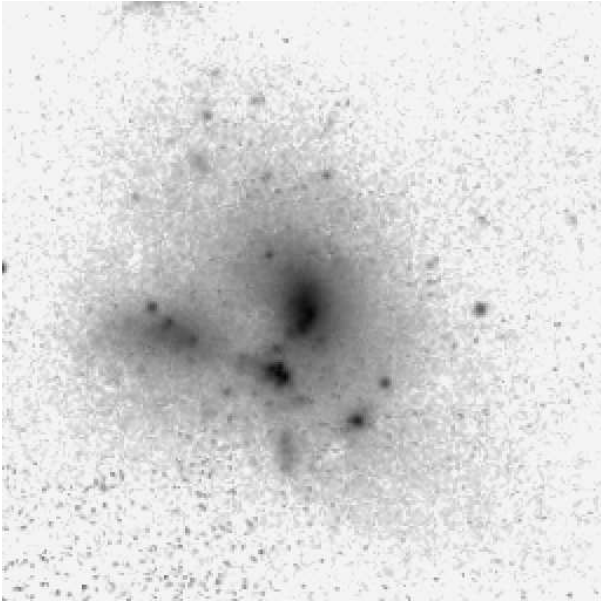}
%	\resizebox{\hsize}{8cm}{\includegraphics{H3242f16.ps}}
      \caption[]{NGC 1487 in H filter. The size of the field is 2.2'x2.2'. 
      North is up, east is to the left. ESO/MPI 2.2-m telescope, La Silla.}
         \label{N1487H}
   \end{figure}

As an example we show in Fig.~\ref{N1487B} and \ref{N1487H} the merger
\object{NGC 1487} in the B and H bands. The B image was obtained with 
the equipment described by Bergvall \& Johansson (\cite{bergvall1}) and the
H image was obtained in August 1993 with ESOs IRAC2 near-IR camera equipped
with a 256x256 NICMOS detector. While the optical image shows the chaotic
signature of a starburst, the near-IR image reveals the smoother structure of
the old stellar component. From a Sky Survey image 
one may get the impression that there are two gas rich galaxies merging 
but it seems more likely from the images that the galaxy is involved in a 
process of a merger between several dwarf galaxies forming a small group.  
There is an increasing observational evidence that this is a common 
phenomenon at high redshifts. E.g. Colina et al. (\cite{colina}) show that 
multiple mergers of sub-L$^*$ galaxies is the dominating process in 
ultraluminous galaxies (ULIGs).

The difference between how we view a merger through the optical window 
versus the near-IR has an impact on our concept of the process in the same 
way as our ignorance of the HI component when we look at images in the 
optical. There is a difference between the result if we merge two massive 
galaxies or if we merge a group of dwarf galaxies even if the final luminosity 
is the same. For one thing the relative gas mass and thus the SFR will be 
higher in the former case, mimicking a starburst. There is also a difference 
in the visual perception that will influence the interpretation. The epoch 
of maximum merger rate probably relates to the mean galaxy density of the 
local universe. Regions of high densities had their most active merger epoch 
at higher redshift while regions of intermediate densities are sites of the 
peak merger activity today. During this epoch there should be an increase in 
relative gas mass fraction compared to 'passive' galaxies since the subunits 
in the mean have a higher gas mass fraction than a single galaxy of equal 
total luminosity. In such case one would expect that the gas mass fraction 
and consequently the SFR is higher than in an isolated galaxy of similar mass. 
The increase in SFR can thus be understood {\it without invoking an increase 
in star formation efficiency (SFE)}.

It is interesting to relate this discussion to some results from CO
observations. Aalto et al. (\cite{susanne1}, \cite{susanne2}, \cite{susanne3} 
and \cite{susanne4}) found unusually large R=$^{12}$CO/$^{13}$CO intensity 
ratios in a sample of luminous mergers. This was interpreted as a consequence 
of turbulent, high-pressure gas in the centres. In a related study, Taniguchi 
\& Ohyama (\cite{taniguchi}) find that both the dust and the $^{13}$CO gas 
are depressed with respect to the $^{12}$CO gas in mergers with high 
FIR luminosities. This would tend to increase the dust temperature if it is 
assumed that the radiation density is unchanged. Mass estimates show however 
that the dust mass is also higher than normal in these galaxies, despite
the depression in dust/gas ratio. The authors conclude that the 
high FIR
luminosity
of the mergers can be explained as a combination of an increased dust
temperature and
a higher dust mass than normal, caused by a high merger frequency of gas and
dust rich
companions. Starbursts that produce
supernova
shocks causing sputtring and grain-grain collisions have been proposed to
explain the
depression of the dust grain content. However, high velocity shocks occur
also
in the
merger event itself and could be a valid alternative and it seems that on
the
whole,
the observations seem to support a scenario with several combined effects
behind
the
high FIR luminosity.

   \begin{figure}
\epsfig{file=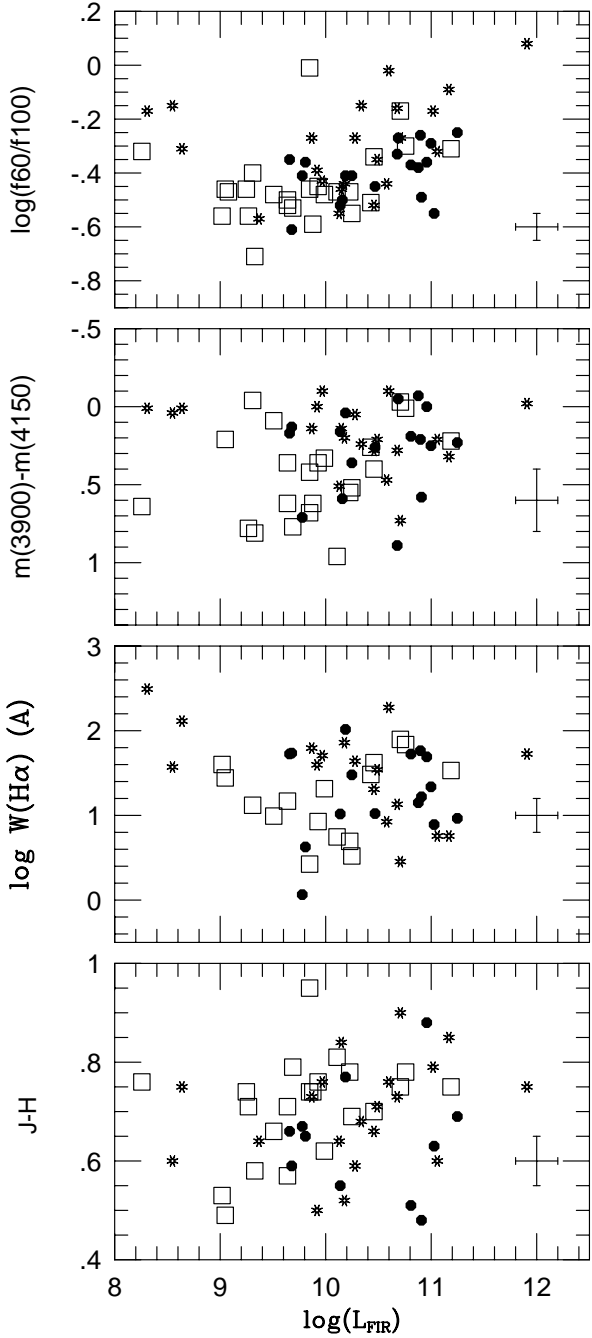, width=8cm}
%	\resizebox{\hsize}{!}{\includegraphics{H3242f17.ps}}
      \caption[]{The correlation between the FIR luminosity and a few
optical/near-IR parameters. The symbols are the same as in Fig. \ref{ubbv}.}
         \label{lfiropt}
   \end{figure}

   \begin{figure}
\epsfig{file=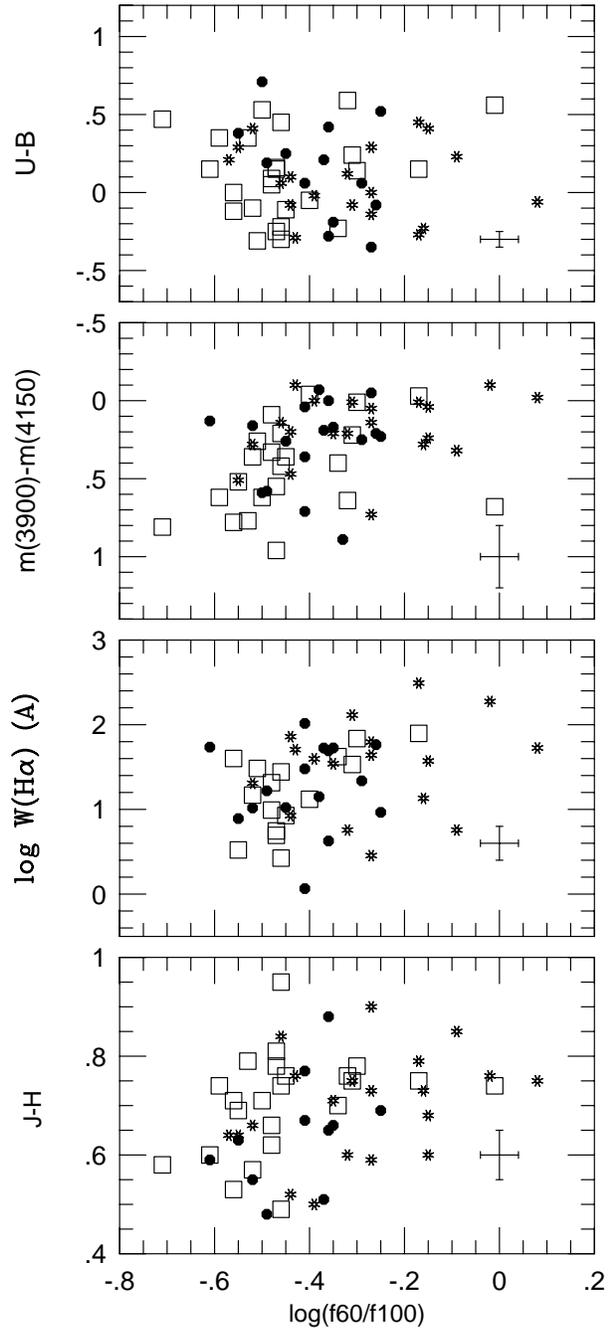, width=8cm}
%	\resizebox{\hsize}{!}{\includegraphics{H3242f18.ps}}
%	\resizebox{\hsize}{12cm}{\includegraphics{tfiropt.ps}}
      \caption[]{The correlation between the FIR temperature index
f$_{60}$/f$_{100}$ and a few optical/near-IR parameters. The symbols
are the same as in Fig. \ref{ubbv}.}
         \label{tfiropt}
   \end{figure}

How do the IRAS data correlate with the optical star
formation signatures? One should keep in mind that we sample regions of
different sizes in different wavelength regions. The IRAS aperture is of the
order of 1 arcminute, i.e. larger than the optical and near-IR apertures and
much larger that the apertures used for the spectroscopy. Fig. 
\ref{lfiropt}
shows the FIR temperature index f$_{60}$/f$_{100}$, the m$_{3900}$-m$_{4150}$
index, \wha and J-H plotted against the FIR luminosity. Except for U-B these
indices are only weakly extinction dependent. We see no strong 
correlation
between these indices and the FIR luminosity in these diagrams but we note
that the two samples are clearly separated in the \wha and the 3900-4150
index
diagrams. Fig. \ref{tfiropt} shows the situation as regards the FIR
temperature, represented by the f$_{60\mu}$/f$_{100\mu}$ ratio. No
correlation
is found with the global U-B colours but significant correlations are found
in the the m$_{3900}$-m$_{4150}$ and \wha diagrams. It seems that warm IRAS
galaxies tend to have more
active star formation in the central regions than colder IRAS galaxies.
This does not necessarily mean that the SFE is higher. 

\begin{figure}
\epsfig{file=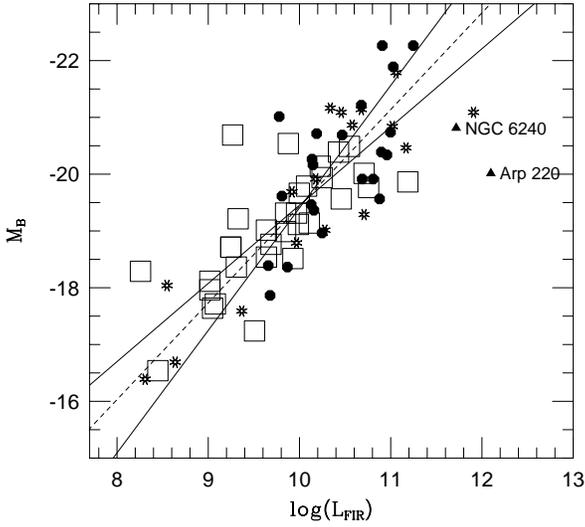, width=8cm}
%	\resizebox{\hsize}{12cm}{\includegraphics{H3242f19.ps}}
      \caption[]{The correlation between absolute magnitude and FIR
luminosity. The symbols are the same as in Fig. \ref{ubbv}. The solid lines 
are the linear regression lines and the dotted line the symmetry line.}
         \label{firm}
   \end{figure}

Fig. \ref{firm} shows the relationship
between the absolute blue magnitude and the FIR luminosity. In order to
improve the statistics, we included an extra set of isolated galaxies
obtained from Bergvall's original list (\cite{bergvall4}). The ESO numbers
of
these galaxies are:
73-09, 74-18, 85-24, 85-47, 143-13, 237-15, 285-20, 287-52, 303-20, 341-23.
These have well determined B magnitudes and IRAS fluxes, all obtained from
the
NED database. In the cases where photometric photometry was not available we
used the total B magnitudes from NED and added a correction term of 0.2
mags. to adapt to the mean difference between our measured photometry and the
total
B magnitude from NED.

We see a strong correlation between B-magnitudes and FIR-fluxes.
It is important to note that the
isolated galaxies follow the same relation as the interacting and merging
galaxies. The interacting galaxies are shifted relative to the isolated ones
with an amount corresponding to a factor of a few in FIR luminosity (cmp.
Fig. \ref{iras}). Thus if the FIR radiation measures star formation
activity, it means that the data indicate an increase in SFR of
the same magnitude, due to the 
interaction. This is quite moderate as compared to the general
notion about induced star formation in such systems. True starbursts are
exceptional and we notice that, in comparison with established massive
starbursts like \object{NGC 6240} and \object{Arp 220} (see positions in the
figure) we find only
one such case in our sample - \object{ESO 286-IG19} (Johansson
\cite{johansson4}).

However, there is an alternative interpretation of the different
distributions
of IGs and
NIGs in figure \ref{firm} that deserves a discussion. The solid lines in the
diagram show the linear regression lines (M$_B$ as function of L$_{FIR}$ and
vice versa)
of the complete data set (excluding
NGC6240 and Arp 220) and the hatched line is the bisector line. The equation
of the bisector line is M$_B$ = -1.703($\pm$0.005) log L$_{FIR}$ -2.41 and
the
correlation coefficient is 0.8. Using a simple model we will show that one
may
predict the value of the slope, -1.7. In the model we will assume that a
star
forming region is a 3-dimensional structure that is opaque in the optical
but
transparent in the far infrared. We will also assume that the mean gas
density
and the SFE is independent of the mass of the gas available
for star formation. Then the IRAS luminosities would simply scale as the
volume V (or the total mass) of the star forming region. The B luminosity,
remembering that the model assumes that the bulk of the blue light
originates
from a thin surface layer, would grow as V$^{2/3}$. We would thus expect 
to
find a relationship $M_B$ = -1.7 log L$_{FIR}$, in perfect agreement with
the
empirical slope. If the SFE increases with luminosity, as would be the 
case
for starbursts, the slope would become flatter.  This is what is observed if one
compares the low luminosity galaxies of our sample with the FIR
starbursts NGC 6240, Arp 220 and ESO 286-IG19, suspected to host starbursts.

We conclude that, at least as concerns the specific sample
discussed here, both the lack of correlation between the FIR and global
optical/near-IR data and the slope of the relation between the optical and FIR
luminosity favour a model where the {\it bulk of the FIR luminosity comes
from regions of normal star formation activity}. The overluminosity in the
IG
sample as compared to the comparison sample may essentially be explained as
a {\it difference in mean mass, relative gas mass fraction, spatial
distribution of gas, stars and dust} between the two groups and the {\it effect 
of shocks in mergers}.
Since mass determinations of interacting
galaxies are often dubious, possible systematic mass differences are
difficult
to check but we show below that the IGs have higher gas masses. A mass
difference is not unexpected, since statistically one would expect to find
the
more massive galaxies in regions where mergers are frequent. These mergers
would involve gas rich low mass galaxies with relatively high SFRs as
compared
to more massive systems and therefore the mean SFR in the IG sample would
automatically be higher than in the NIG sample, disregarding the possible
effects of tidal effects and mergers. We would thus predict that IGs have 1)
higher masses than isolated galaxies and 2) higher relative HI content than
normal galaxies of the same mass.

   \begin{figure}
\epsfig{file=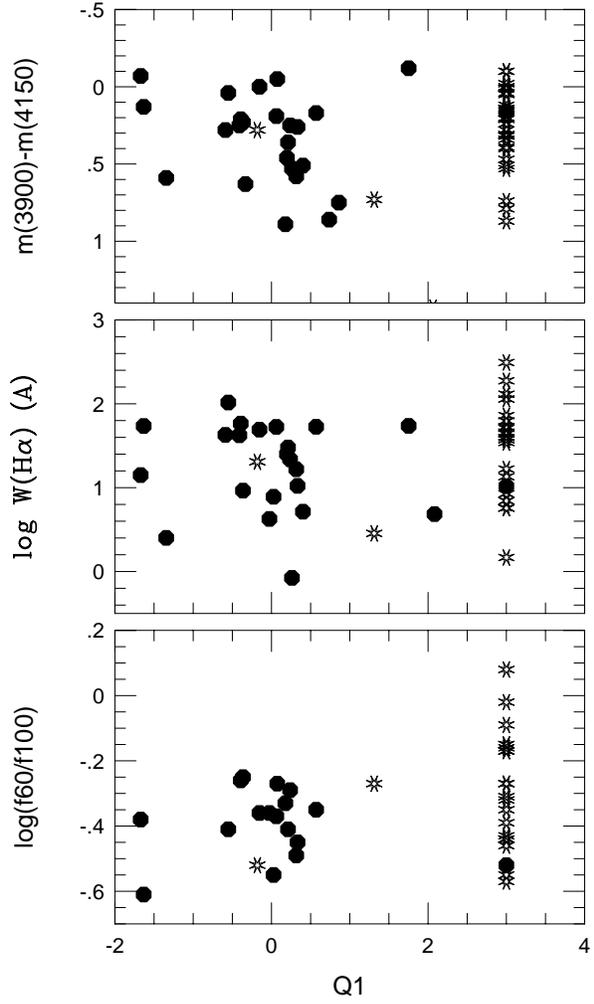, width=8cm}
%	\resizebox{\hsize}{12cm}{\includegraphics{H3242f20.ps}}
      \caption[]{The Q1 parameter (Dahari \& de Robertis \cite{dahari})
versus various star formation parameters for our sample of interacting 
galaxies. Symbols as in Fig. 4. Increasing strength of 
interaction $\rightarrow$ higher Q1 value. No value of the parameter can be 
derived
for most of the merger candidates (stars) which have been placed at Q1=3 for 
comparison. }
         \label{q1}
   \end{figure}

   \begin{figure}
\epsfig{file=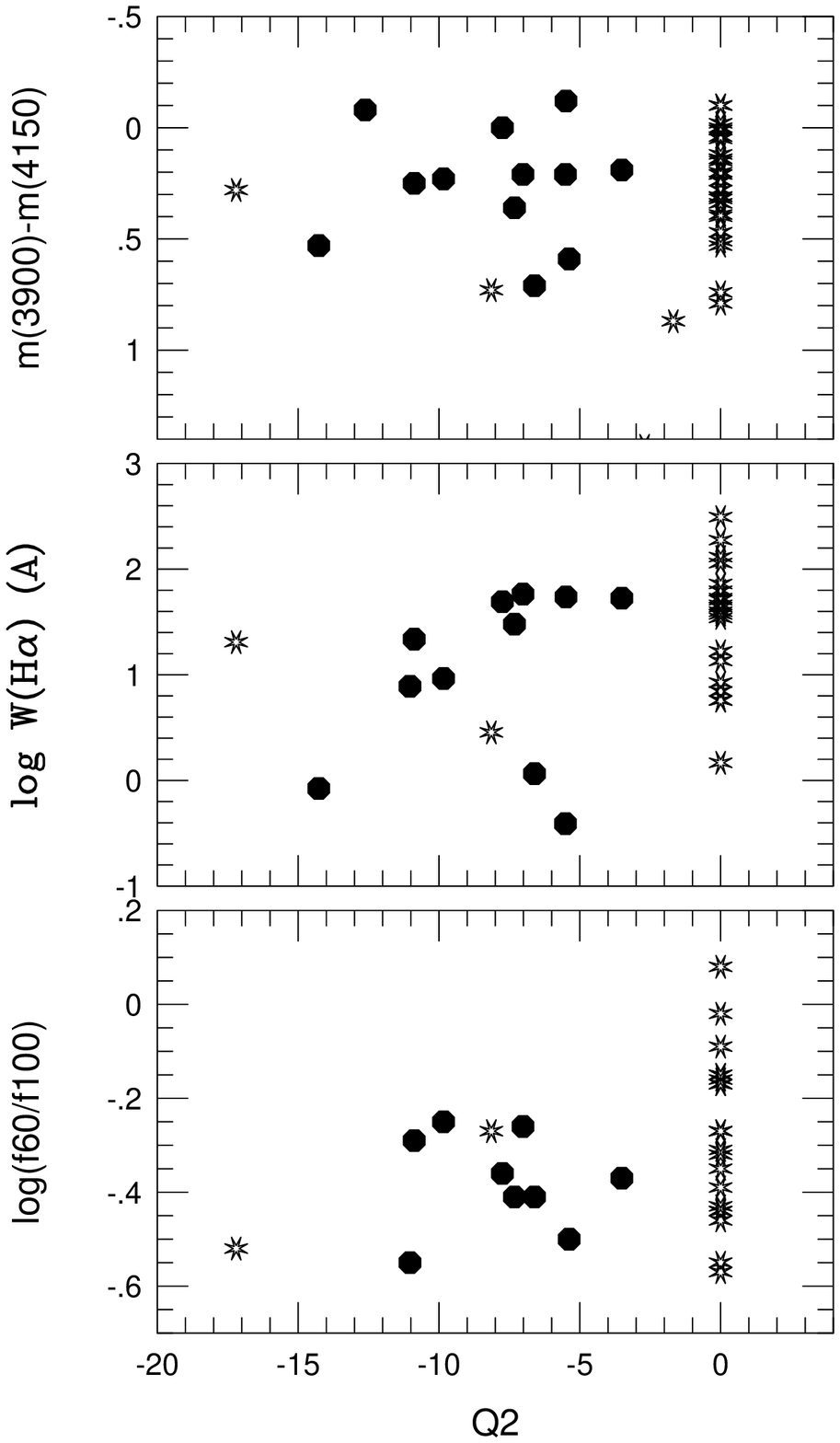, width=8cm}
%	\resizebox{\hsize}{12cm}{\includegraphics{H3242f21.ps}}
      \caption[]{The Q2 parameter (Dahari \& de Robertis \cite{dahari})
versus various star formation parameters for our sample of interacting 
galaxies. Symbols as in Fig. 4. Increasing strength of 
interaction $\rightarrow$ higher Q2 value. No value of the parameter can be 
derived
for most of the merger candidates (stars) which have been placed at Q2=0 for 
comparison.}
         \label{q2}
   \end{figure}

\subsection{Neutral gas}

CO-observations of part of our sample were obtained by Horellou \& Booth
(\cite{horell1}, \cite{horell2}). The results can in principle be interpreted in
terms of increased SFR in the IG sample, because the SFE, measured as
$L_{FIR}$/\m(H$_2$), is somewhat higher for them. However, this difference
may
as
well be related to different distributions of stars and dust. Solomon \&
Sage
(\cite{solomon}) also argue that the observed high SFE in late interacting
systems may arise from an underestimation of the H$_2$ mass in these
galaxies
by the CO luminosity.

   \begin{figure}
\epsfig{file=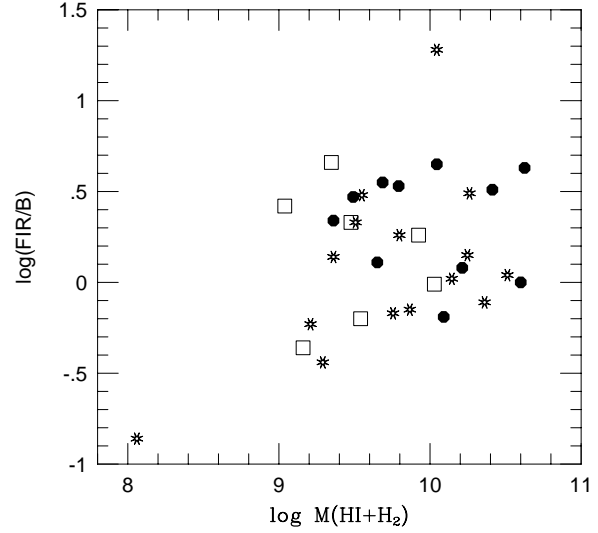, width=8cm}
%	\resizebox{\hsize}{!}{\includegraphics{H3242f22.ps}}
      \caption[]{The relation between total gas mass and FIR/B luminosity 
      ratio for the sample galaxies. }
         \label{h1firb}
   \end{figure}

\begin{figure}
\epsfig{file=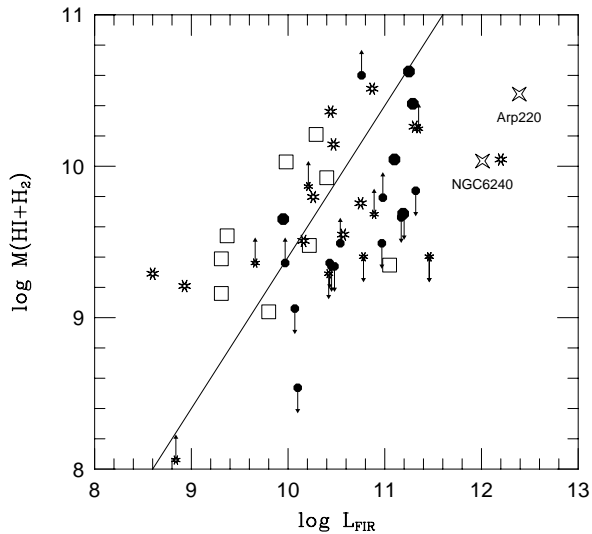, width=8cm}
%	\resizebox{\hsize}{!}{\includegraphics{H3242f23.ps}}
      \caption[]{The relation between total gas mass and FIR luminosity for
the sample galaxies. Upper and lower limits are marked with arrows. The solid
line shows the slope of the \m (HI+H$_2) \propto$ L$_{FIR}$, arbitrarily shifted
to fit the data.}
         \label{h1fir}
   \end{figure}

If the SFE increases as a consequence of the interaction, one would expect
L$_{FIR}$/L$_B$ to increase with gas mass. There is no such trend 
(Fig.~\ref{h1firb}) for those galaxies in our samples that
have
measured HI and/or H$_2$ masses. Most of the data for the IG sample were
obtained from Horellou \& Booth (\cite{horell1}). Data for the rest of the
galaxies were obtained from NED. When only HI masses were available,
corrections to convert these data to total hydrogen gas masses were done,
using the morphological classifications and the relationship between
morphological types and the \hm/\2hm mass ratio from Young \& Scoville
(\cite{young3}). Figure \ref{h1fir} shows the gas
masses as a function of L$_{FIR}$.  Although the data are scarce, two 
things 
are evident from
the
figure: 1) The hydrogen masses are higher for the IGs with roughly by a factor
of 2-3; 2) As seen from the
shape of the upper envelope of the distribution, there is roughly a linear
relationship between \m (HI+H$_2$) and L$_{FIR}$. Both these results agree with 
a model
in which the differences in the IRAS data between the samples to a
significant amount are due to differences in gas mass. The statistics 
are 
poor, but if we assume that there is a difference in gas mass of a factor of 2-3 
and also remember that the IGs in the mean have higher FIR temperatures, that 
will boost the total FIR emission, the difference in FIR luminosity can be fully 
accounted for.  The three probable massive
starbursts Arp 220, NGC 6240 and ESO 286-IG19 clearly deviate from the rest
of
the sample. The estimated star formation rates in these three galaxies, of
the
order of 100 \sma yr$^{-1}$ or more (Kennicutt \cite{kennicutt4}, Johansson
\cite{johansson2}), are indeed outstanding.

\section{Tidal action}

Part of the dispersion for interacting galaxies in
the color-color diagram may also be related to different orbital geometries
of the encountering galaxies. The importance of orbital geometry on
reactivity
of galactic disks was addressed already by Toomre \& Toomre (\cite{toomre1}). 
The star
formation properties of the encountering galaxies depend, among other
things,
on the orbital inclination, on the sense of the spinning of the disks and
the
mass distribution in the galaxies. The effects of orbital inclination on
star
formation has been discussed for example by Salo \& Laurikainen 
(\cite{salo2},
\cite{salo3}) for the pairs NGC 7753/7752 (Arp 86) and M51. They
demonstrated
that in NGC 7753/7752,
where the companion moves nearly in the main galactic disk mass transfer to
the companion is possible, whereas in M51, for the nearly perpendicular
orbit
of the companion it is not allowed.

It would be interesting to investigate in more detail how interaction
affects
star formation both in the disk and in the centre. At this stage we will
only
discuss the possible correlation between the strength of the tidal
interaction
between the two components of the system and the parameters already
discussed, \xwha, m$_{3900}$-m$_{4150}$ and the IRAS temperature. As a rough
measure of the strength of interaction we will use the Q$_1$ and Q$_2$
parameters, defined by Dahari \& de Robertis (\cite{dahari}). The
interesting result (Fig.~\ref{q1} and Fig.~\ref{q2}) is that none of the 
star forming parameters show any
strong
correlation. Only dust temperature shows a weak rather insignificant
correlation in Fig.~\ref{q1} and only the mergers show high
temeratures. This could be understood as an effect of a redistribution of the
dust towards the centre, probably leading to a more efficient heating of the
dust by the old stars (e.g. \cite{thronson}).  It means that, {\it on
average},
the star formation seen in the IGs is not much ruled by the strength of the
interaction. The orbital parameters have been shown to be important for mass
transfer under certain conditions (e.g. as in NGC 7753/7752) but these
events
are rare and the induced events of star formation may be shortlived.
Conditions
for true starbursts may be more frequent in mergers and related to their
internal
structure rather than the orbital parameters of the preceding close
encounters
(Mihos \& Hernquist \cite{mihos}). The total mass and relative gas content
are
other important parameters.

\section{Discussion}

As mentioned above, there is only one true massive starburst galaxy in
the sample. Our IG sample constitutes of 8\% of the total number of
galaxies in the ESO/Uppsala catalogue at the same magnitude limit.
If our sample is representative of galaxies in general, it also means that
about 0.1\% of a magnitude limited sample are massive starbursts generated 
by interactions and mergers. On the other hand virtually all luminous infrared 
galaxies (ULIRGS) have distorted morphologies (Joseph \& Wright \cite{joseph}, 
Armus et al. \cite{armus}, Sanders et al. \cite{sanders2}, Clements et al. 
\cite{clements}, Murphy et al. \cite{murphy}), indicating recent mergers. This 
leaves little room for other mechanisms that could trigger massive starbursts. 
Therefore the figure is likely to hold also for the total population of such 
cases. Such rare
phenomena have no important influence on the mean star formation
rate of galaxies in general. This conclusion agrees qualitatively with that
of Kennicutt
et al. (\cite{kennicutt2}) although they claim a higher rate of "active
galaxies". It also agrees with the result from Allam`s study of IRAS data of 
interacting galaxies (Allam \cite{allam}) and is supported by the lack of 
influence of
interactions on the ratio between the FIR and the 2.4 Ghz radio emission
found by Lisenfelt et al. (\cite{lisenfeld2}).

But what is the situation like at higher
redshifts? The evolution of the merger rate with redshift is apparently an
important parameter. The results from recent studies are somewhat
contradictive. From model comparisons with ISO and optical data Roche and
Eales (\cite{roche}) find support for a strong evolution to z$\sim$1 and
estimate the merger rate to increase with redshift as (z+1)$^2$. The large
observational redshift survey projects, however, seem to show no evolution
(Carlberg et al. \cite{carlberg}) in this redshift interval. Infante et al.
(\cite{infante}), working at slightly lower redshifts, are somewhere in
between. At even higher redshifts there are no strong indications 
of a 
further increase in merger rate of luminous galaxies. Thus, for the upper part 
of the LF one would not expect more than a few times increase in merger rate to 
z$\sim$ 1.5 and then insignificant changes towards higher redshifts. If massive 
starbursts increase at the same rate, it would mean that
less than a few times 0.1\% of a magnitude limited sample of galaxies would be
involved in massive starbursts. Since the SFR in such cases is 
enhanced with 
at least one
magnitude it could be of significance for the mean gas consumption rate but
for a typical galaxy, most stars would form in a more quiescent phase.

If tidally induced massive starbursts would dominate the star formation
process at z=1 most of the metals would also be formed at this epoch since 
the gas consumption time scale would be of the order of a few 10$^8$ yr. 
From recent observations it is quite evident that this is not the case. Star 
formation activity at higher redshifts
can also be studied directly. The largest sample of spectroscopically
confirmed high redshift galaxies selected by narrow band infrared imaging
was investigated by Moorwood et al. (\cite{moorwood}). They derived star
formation rates from the \ha fluxes. The peak rates they found were in the
range 20-35 \sma yr$^{-1}$ without extinction correction (but the
observations
indicate low extinction). The velocity dispersions implied masses around
10$^{10}$\sma. The gas consumption timescale is thus of the order of the age
of
the
universe at that redshift. These objects should not be labeled
starburst galaxies. It seems more likely that these galaxies, being the
brightest in the sample, are ellipticals in the formation phase exposed to
continuous infall of gas clouds and dwarf galaxies.

Recent SCUBA results (Lilly et al. \cite{lilly}) open the possibility for a
significant portion of hidden massive starbursts but the interpretation of
the
infrared data is not univocal. Possibly, and maybe a major part of the
emission is powered by AGNs (Lilly et al. \cite{lilly}). From these data one
cannot reach any conclusion
regarding if starbursts of the Arp 220 type have a significant influence on
the gas consumption and metal production at high redshifts.

\section{Conclusions}

We have discussed the star formation properties of a magnitude limited
sample of interacting and merging galaxies (IGs) in comparison with a sample
of isolated galaxies (NIGs) of similar morphological types. In particular we
have compared a number of star formation signatures. The {\it global}
optical colours {\it do not} support a significant increase in star formation
activity in the IG sample. On the other hand the colours of the central regions
differ which can be understood as a combination of extinction effects and a 
larger proportion of young stars. The \ha equivalent widths of the
centres (although hampered by small number statistics), are 2-3 times higher
in the IG sample indicating a modest increase in star formation activity in
these regions. However, equivalent widths characteristic of true starbursts
are rare. Starbursts do occur in our sample but are very uncommon
or shortlived. In the general case we do not find any strong support that 
interactions and mergers trigger true starbursts. This is in conflict with the 
generally accepted view founded by Larson \& Tinsley (\cite{larson2}), in which 
interactions are causing significant changes in global broadband colours, 
typical of an increasing contribution from young stars. We show that this 
result, based on a comparison between Arp galaxies and normal Hubble type 
galaxies, is probably an artifact caused by an underestimate of the errors in 
the photometry of the Arp sample.

We find that the interacting and merging galaxies in comparison with isolated 
galaxies are more luminous in optical and near-IR and that the dust temperature 
is higher. While this often is claimed as an effect of starburst activity, it 
may also be explained if the IGs in general have higher masses than the NIGs 
and/or the gas, dust and young stars are more centrally concentrated. The latter 
could also explain the increase in the \ha equivalent width in the central 
regions. Mass transport to the central region also causes mechanical heating and 
shock ionization. An excess in mass means higher mean metallicity and dust 
content, further increasing the FIR luminosity. Since mass determinations of 
dynamically distorted systems are difficult to carry out, such a situation may 
have been overlooked in many cases.

Many of the previous investigations on this issue, claiming frequent
occurrences of tidally triggered starbursts, are hampered by strong selection
effects, often biased towards IGs with infrared excess. Although we worry 
about the selection effects we want to emphasize that we do not claim that 
starbursts are {\it not} triggered by interactions or 
mergers. On the contrary, mergers seem to be a necessity to create 
true starbursts like those in blue compact galaxies or some ULIRGs. But it does 
not appear to be a sufficient condition. The role of dark matter, angular 
momentum 
flow and star formation processes in general need to be understood 
better. Our estimate is that about 0.1\% of all galaxies in a magnitude
limited sample are true starbursts. Even taking into account the increase in
merger rate with redshift we find that tidal interaction and mergers probably
do not have a major effect on the gas consumption rate at these epochs. The
observational data are yet too meager for final conclusions but if confirmed, it
will have a significant impact on theoretical modelling of early galaxy
evolution (e.g. Bekki \cite{bekki}; Devriendt \& Guiderdoni \cite{devriendt};
Guiderdoni et al. \cite{bruno}). Obvious consequences are that there is no
cosmic epoch at redshifts z$<$5 at which the mean gas consumption time scales in
massive galaxies are significantly shorter than the Hubble age at that epoch.
Moreover, the effects of superwinds created by coeval supernova blasts would
have less effects on the morphology of the massive galaxies and that it
would be more difficult to reionize the universe at high redshifts because the
ionization parameter would be reduced.

\begin{acknowledgements}

We are indebted to Lennart Johansson for his contributions in the early
phase of this project. Our referees are thanked for many useful comments that 
have helped to improve the quality of this paper. N. Bergvall gratefully 
acknowledges partial 
support
from  the Swedish Natural Science Research Council. We thank the ESO staff
for helpful assistance. This research has made use of the NASA/IPAC
Extragalactic Database (NED) which is operated by the Jet Propulsion
Laboratory,
California Institute of Technology, under contract with the National
Aeronautics
and Space Administration.
We have made use of the LEDA database (http://leda.univ-lyon1.fr).

\end{acknowledgements}


\begin{thebibliography}{}

\bibitem[1991]{susanne1}Aalto, S., Black, J. H., Johansson, L. E. B., Booth,
R.S., 1991, A\&A, 249, 323
\bibitem[1995]{susanne2}Aalto, S., Booth, R. S., Black, J. H., Johansson, L.
E.B., 1995, A\&A, 300, 369
\bibitem[1997]{susanne3}Aalto, S., Radford, Simon J. E., Scoville, N. Z.,
Sargent, A. I, 1997, ApJ 475, L107
\bibitem[2000]{susanne4}Aalto, S., H\"uttemeister, S., 2000, A\&A 362, 42
\bibitem[1996]{abraham}Abraham, R. G., Tanvir, N. R., Santiago, B. X.,
Ellis, R. S., Glazebrook, K., van den Berg, S., 1996, MNRAS 279, L47
\bibitem[1976]{alcaino}Alcaino, G., 1976, A\&AS 26, 261
\bibitem[1996]{allam}Allam, S, 1996, "IRAS study of interacting galaxies", 
Ph.D. Thesis, Potsdam
\bibitem[1987]{appleton}Appleton, P.N., Struck-Marcell, C., 1987, ApJ 312, 556
\bibitem[1987]{armus}Armus, L., Heckman, T. M., Miley, G. H., 1987, AJ 94, 831
\bibitem[1966]{arp}Arp, H.C., 1966, Atlas of Peculiar Galaxies, Pasadena,
California Institute of Technology
\bibitem[1987]{arp2}Arp, H. C., Madore, B., 1987, "A catalogue of southern
peculiar
galaxies and associations", Cambridge, New York : Cambridge University Press
\bibitem[1997]{bachall}Bachall J. N., Kirhakos S., Saxe D. H., Schneider D.
P., 1997, ApJ 479, 642
\bibitem[1990]{barnes2}Barnes, J.: 1990, in 'Dynamics and Interactions of
Galaxies', ed. R. Wielen, Springer, Berlin Heidelberg
\bibitem[1991]{barnes3}Barnes, J. E., Hernquist, L. 1991, ApJ 370, L65
\bibitem[2000]{barton}Barton, Elizabeth J., Geller, Margaret J.,
 Kenyon, Scott J., 2000, ApJ 530, 660
\bibitem[2001]{bekki}Bekki, K., 2001, ApJ 546, 189
\bibitem[2000]{bekki2}Bekki, K., Shioya, Y., 2000, PASJ 52, L57
\bibitem[1978]{bergvall5}Bergvall, N. A. S., Ekman, A. B. G., Lauberts, A.;
Westerlund, B. E., Borchkhadze, T. M., Breysacher, J., Laustsen, S., Muller, A.
B., Schuster, H. E., Surdej, J., West, R. M., 1978, A\&AS 33, 243
\bibitem[1981a]{bergvall3}Bergvall, N., 1981, Uppsala Astronomical 
Observatory Report No. 18
\bibitem[1981b]{bergvall4}Bergvall, N., 1981, Uppsala Astronomical 
Observatory Report No. 19
\bibitem[1995]{bergvall1}Bergvall, N., Johansson, L., 1995, A\&AS 113, 499
\bibitem[1989]{bergvall2}Bergvall, N., R\" onnback, J, Johansson, L., 1989,
A\&A 222, 49
\bibitem[1984]{burstein}Burstein, D., Heiles, C., 1984, ApJS 54, 33
\bibitem[1986]{bushouse2}Bushouse, H. A. 1986, AJ 91, 255
\bibitem[1988]{bushouse1}Bushouse, H. A., Lamb V. A., Werner, M. W. 1988,
ApJ 335, 74
\bibitem[2000]{carlberg}Carlberg, R. G., Cohen, J. G., Patton, D. R.,
Blandford, R., Hogg, D. W., Yee, H. K. C., Morris, S. L., Lin, H., Hall,
Patrick B., Sawicki, M., Wirth, Gregory D., Cowie, L. L., Hu, E., Songaila,
A., 2000, ApJ 532,1
\bibitem[2001]{carter}Carter, B. J., Fabricant, D. G., Geller, M. J., 
Kurtz, M. J., McLean, B., 2001 ApJ 559, 606
\bibitem[1984]{chincarini}Chincarini, G., Tarenghi, M., Sol, H., Crane, P.,
Manousoyannaki, I., Materne, J., 1984, A\&ApS 57, 1
\bibitem[1996]{clements}Clements, D. L., Sutherland, W. J., McMahon, R. G., 
Saunders, W., 1996, MNRAS 279, 477 
\bibitem[2001]{colina}Colina, Luis, Borne, Kirk, Bushouse, Howard, Lucas, 
Ray A., Rowan-Robinson, Michael, Lawrence, Andy, Clements, David, Baker, 
Amanda, Oliver, Seb, 2001, ApJ 563, 546
\bibitem[1984]{dahari1}Dahari, O., 1984, AJ 89, 966
\bibitem[1985]{dahari2}Dahari, O., 1985, AJ 90, 1772
\bibitem[1988]{dahari}Dahari, O., de Robertis, M.M., 1988, ApJ Suppl. Ser. 67, 
249
\bibitem[1998]{de robertis}de Robertis, M. M., Yee, H. K. C., Hayhoe, K.,
1998, ApJ 496, 93
\bibitem[2000]{devriendt}Devriendt, J.E.G., Guiderdoni, B.,
2000, A\&A, 263, 851
\bibitem[1997]{donzelli}Donzelli, C.J, Pastoriza, M.G, 1997, ApJS 111, 181
\bibitem[1999]{deborah}Dultzin-Hacyan, D., Krongold, Y., Fuentes-Guridi,
I., Marziani, P., 1999, ApJ 513, L111
\bibitem[1997]{fioc1}Fioc, M., Rocca-Volmerange, B., 1997, A\&A 326, 950
\bibitem[2000]{fioc2}Fioc, M., Rocca-Volmerange, B., 2000, astro-ph/9912179
\bibitem[1998]{franceschini}Franceschini, A., Silva, L., Fasano, G.,
Granato, L., Bressan, A., Arnouts, S., Danese, L., 1998, ApJ 506, 600
\bibitem[1987]{frenk}Frenk, C.S., White, S.D.M., Davis, M., Efstatiou, G.:
1987, ApJ 327, 507
\bibitem[1991]{gavazzi}Gavazzi, Giuseppe, Boselli, Alessandro,
Kennicutt, Robert, 1991, AJ 101, 1207
\bibitem[1980]{griersmith}Griersmith, D., 1980, MNRAS 191, 1
\bibitem[1998]{bruno}Guiderdoni, B., Hivon, E., Bouchet, F.R., Maffei, B.,
1998, MNRAS, 295, 877
\bibitem[1990]{heckman}Heckman, T. M., Armus, L., Miley, George K., 
1990, ApJS 74, 833
\bibitem[1989]{hernquist2}Hernquist, L, 1989, Nature 340, 687
\bibitem[1997]{ho}Ho, L. C., Filippenko, A. V., \& Sargent, W. L. W., 
1997, ApJ 487, 568
\bibitem[1980]{holmberg}Holmberg, E., Lauberts, A., Schuster, H.-E., 
West, R.M.: 1980, A\&AS 39, 173
\bibitem[1997]{horell1}Horellou, C., Booth, R., A\&AS 126, 3
\bibitem[1999]{horell2}Horellou, C., Booth, R., Karlsson, B., 1999, 
Ap\&SS 269/270, 629
\bibitem[1980]{hummel}Hummel, E., 1980, A\&A 89, L1
\bibitem[1990]{hummel2}Hummel, E., van der Hulst, J. M., Kennicutt, R. C.,
Keel, W. C., 1990, A\&A 236, 333
\bibitem[1996]{infante}Infante L., de Mello D. F., Menanteau F., 1996, ApJ,
469, L85
\bibitem[2000]{jansen}Jansen, Rolf A., Franx, Marijn, Fabricant, Daniel,
 Caldwell, Nelson, 2000, ApJS, 126, 271
\bibitem[1988]{johansson1}Johansson, L., 1988, A\&A 191, 29
\bibitem[1990]{johansson2}Johansson, L., 1990, A\&A, 241, 389
\bibitem[1991]{johansson4}Johansson, L., 1991, A\&A 241, 38
\bibitem[1990]{johansson3}Johansson, L., Bergvall, N., 1990, A\&AS 86, 167
\bibitem[1985]{joseph}Joseph, R. D., Wright, G. S. 1985, MNRAS, 214, 
87
\bibitem[1973]{kara1}Karachentseva, V. E., 1973, Soobshch. Spets. Astrof. 
Obs., 8, 3
\bibitem[1997]{kara2}Karachentseva, V. E., Lebedev, V. S., Shcherbanovskij, A. 
L., 1997, VizieR On-line Data Catalog: VII/82A. 
\bibitem[1983]{keel2}Keel, W., 1983, ApJS, 52, 229
\bibitem[1985]{keel}Keel, W.C., Kennicutt, Jr., R.C., Hummel, E., van der
Hulst, J.M., 1985, AJ 90, 708
\bibitem[1984]{kennicutt1}Kennicutt, Jr., R.C., Keel, W.C., 1984, ApJ 279,
L5
\bibitem[1987]{kennicutt2}Kennicutt, Jr. R. C., Keel, W. C., van der Hulst,
J. M., Hummel, E., Roettiger, K. A., 1987, AJ 93, 1011
\bibitem[1998]{kennicutt4}Kennicutt, Jr., R.C., 1998, ApJ 498, 541
\bibitem[1998]{kennicutt3}Kennicutt, Jr., R.C., 1998, in "Galaxies,
Interactions and Induced Star Formation", ed. Jr. R. C. Kennicutt, F.
Schweitzer
and J. E. Barnes, Saas-Fee, Springer
\bibitem[1990]{kormendy}Kormendy, J.: 1990, in 'Dynamics and Interactions of
Galaxies', ed. R. Wielen, Springer, Berlin Heidelberg
\bibitem[2000]{kunth}Kunth, D., \"Ostlin, G., 2000, A\&ARv, 10, 1
\bibitem[1993]{lacey}Lacey, C., Guiderdoni, B., Rocca-Volmerange, B., Silk,
J., 1993, ApJ 402, 15
\bibitem[1978]{larson2}Larson, R.B., Tinsley, B.,M., 1978, ApJ 219, 46
\bibitem[1984]{lauberts}Lauberts, A., 1984, A\&AS 58, 249
\bibitem[1982]{lauberts2}Lauberts, A., 1982, "The ESO/Uppsala Survey of the 
ESO(B) Atlas",  ESO
\bibitem[1995]{laurik2}Laurikainen, E., Salo, H., 1995, A\&A 293, 683
\bibitem[1989]{laurik3}Laurikainen, E., Moles, M., 1989, ApJ 345, 176
\bibitem[2000]{lefevre}Le F\`evre, O., Abraham, R., Lilly, S. J.,
Ellis, R. S., Brinchmann, J., Schade, D., Tresse, L., Colless, M.,
Crampton, D., Glazebrook, K., Hammer, F., Broadhurst, T., 2000, MNRAS 311, 565
\bibitem[1999]{lilly}Lilly, S.J., Eales, S.A., Gear, W.K.P., Hammer, F., Le
F\'evre, O., Crampton,  D., Bond, R.J., Dunne, L., 1999, ApJ 518, 641
\bibitem[1996]{lisenfeld2}Lisenfeld, U., V\"olk, H.J., Xu, C., 1996, A\&A 314, 
745
\bibitem[2000]{lisenfeld}Lisenfeld, U., Isaak, K. G., Hills, R., 2000, MNRAS 
312, 433
\bibitem[1995]{liu}Liu, C.T., Kennicutt, R.C., 1995, ApJ 450, 547
\bibitem[1985]{lonsdale}Lonsdale, C.J., Helou, G., 1985, 'Catalogued galaxies
and quasars observed in the IRAS survey', Pasadena: JPL
\bibitem[1970]{mathis}Mathis, J.,1970, ApJ 159, 263
\bibitem[1989]{mckenty}McKenty, J.W., 1989, ApJ, 343, 125
\bibitem[1994a]{mcleod1}McLeod, K. K., Rieke, G. H., 1994, ApJ 431, 137
\bibitem[1994b]{mcleod2}McLeod, K.K, Rieke, G.H., 1994, ApJ 420, 58
\bibitem[1994]{mihos}Mihos, J. C., Hernquist, L., 1994, ApJ 425, 13
\bibitem[1993]{mobasher}Mobasher, B., Sharples, R. M., Ellis, R. S., 1993,
MNRAS 263, 560
\bibitem[2000]{moorwood}Moorwood, A. F. M., van der Werf, P. P., Cuby, J.
G., Oliva, E., 2000, A\&A 362, 9
\bibitem[1990]{moshir}Moshir, M. et al., 1990, Infrared Astronomical
Satellite Catalogs - The Faint Source Catalog, Version 2.0
\bibitem[1996]{murphy}Murphy, T. W., Armus, L., Matthews, K., Soifer, B. T., 
Mazzarella, J. M., Neugebauer, G., 1996, AJ 111, 102
\bibitem[1988]{noguchi2}Noguchi, M. 1988, A\&A 203, 259
\bibitem[1991]{osterbrock}Osterbrock, D., 1991, PASP 103, 874
\bibitem[1997]{patton}Patton, D. R., Pritchet, C. J., Yee, H. K. C.,
Ellingson, E., Carlberg, R. G., 1997, ApJ 475, 29
\bibitem[1982]{peterson2}Peterson, C.J., 1982, PASP 94, 404
\bibitem[1986]{peterson}Peterson, C.J., 1986, PASP 98, 1273
\bibitem[1984]{quinn}Quinn, P.J., 1984, ApJ 279, 596
\bibitem[1999]{ramella}Ramella, M., Zamorani, G., Zucca, E., Stirpe, G. M., 
Vettolani, G., Balkowski, C., Blanchard, A., Cappi, A., Cayatte, V., 
Chincarini, G., Collins, C., Guzzo, L., MacGillivray, H., Maccagni, D., 
Maurogordato, S., Merighi, R., Mignoli, M., Pisani, A., Proust, D., 
Scaramella, R, 1999, A\&A 342,1
\bibitem[1990]{rocca}Rocca-Volmerange, B., Guiderdoni, B.: 1990, A\&A 227, 362
\bibitem[1999]{roche}Roche, N., Eales, S.A., 1999, MNRAS 307, 111
\bibitem[1982]{sadler}Sadler, E., 1982, Ph.D. thesis, Mt Stromlo and Siding
Spring observatories
\bibitem[1991]{salo1}Salo, H., 1991, A\&A 243, 118
\bibitem[1993]{salo2}Salo, H., Laurikainen E., 1993, ApJ 410, 586
\bibitem[2000]{salo3}Salo, H., Laurikainen E., 2000, MNRAS 319, 377
\bibitem[1961]{sandage}Sandage, A., 1961, The Hubble Atlas of Galaxies
(Washington: Carnegie Institution of Washington)
\bibitem[1978]{sandage2}Sandage, A., Visvanathan, N., 1978 ApJ 223, 707
\bibitem[1996]{sanders}Sanders, D.B, Mirabel, I.F., 1996, Ann Rev Astr. Ap 34,
749
\bibitem[1988]{sanders2}Sanders, D. B., Soifer, B. T., Elias, J. H., Madore, B. 
F., Matthews, K., Neugebauer, G., Scoville, N. Z., 1988, ApJ, 325, 74
\bibitem[1986]{scalo}Scalo, J.M., 1986, Fund. Cosmic Phys. 11, 1
\bibitem[1966]{shobbrook}Shobbrook, R.R., 1966, MNRAS 131, 351
\bibitem[1980]{schweizer1}Schweizer, F., 1980, ApJ 237, 303
\bibitem[1990]{schweizer2}Schweizer, F., 1990, in 'Dynamics and Interactions
of Galaxies', ed. R. Wielen, Springer, Berlin Heidelberg
\bibitem[1992]{schweizer3}Schweizer, F., Seitzer, P., 1992, AJ 104, 1039
\bibitem[1999]{schweizer5}Schweizer, F., 1999, in Physical Reports 321, No 1-3
\bibitem[1992]{sekig}Sekiguchi, K., Wolstencroft, R. D., 1992, MNRAS 255, 58
\bibitem[1988]{solomon}Solomon, P.M, Sage, L.J., 1988, ApJ 334, 613
\bibitem[1995]{spinoglio}Spinoglio, L., Malkan, M.A., Rush, B.,
Carrasco, L., Recillas-Cruz, E., 1995, ApJ 453, 616
\bibitem[1982]{stauffer}Stauffer J., 1982, ApJS, 50, 517
\bibitem[1998]{taniguchi}Taniguchi, Y., Ohyama, Y., 1998, ApJ 508, L13
\bibitem[1988]{telesco}Telesco, C. M., Wolstencroft, R. D., Done, C., 1988, ApJ 
329, 174
\bibitem[1991]{terlevich}Terlevich, R., Melnick, J., Masegosa, J., Moles, M.,
Copetti, M. V. F., 1991, A\&AS 91, 285
\bibitem[1990]{thronson}Thronson, H.A., Jr., Majewski, S., Descartes, L., 
Hereld, M., 1990, ApJ 364, 456
\bibitem[1972]{toomre1}Toomre, A., Toomre, J. 1972, ApJ 178, 623
\bibitem[1977]{toomre2}Toomre, A., 1977, in ``The Evolution of Galaxies and
Stellar Populations, ed. B. Tinsley, R. Larson, New Haven, Yale University
Obs., p. 401
\bibitem[1990]{vdb}van den Bergh, S., 1990, QJRAS 31, 153
\bibitem[1976]{dev}de Vaucouleurs, G., de Vaucouleurs, A., Corwin J.R.,
1976, Second Reference Catalogue of Bright Galaxies, Austin: University of
Texas Press
\bibitem[1977]{dev3}de Vaucouleurs, G., in "The Evolution of Galaxies and
Stellar Populations", Tinsley, B.M. and Larson, R.B. (eds), Yale Univ. Press
\bibitem[1991]{dev4}de Vaucouleurs, G., de Vaucouleurs, A., Corwin, J. R., 
Buta, R. J., Paturel, G., Fouque, P., 1991, "Third reference catalogue of 
Bright galaxies", New York : Springer-Verlag
\bibitem[1981]{west}West, R. M., Surdej, J., Schuster, H.-E., Muller, A. B.,
Laustsen, S., Borchkhadze, T. M, 1981, A\&AS 46, 57
\bibitem[1979]{white}White, S., 1979, Monthly Notices Roy. Astron. Soc. 189, 831
\bibitem[1997]{whitemore}Whitemore, B. C., Miller, B. W., Schweizer, F.,
Fall, S. M., 1997, AnJ 114, 1797
\bibitem[1992]{witt}Witt, A. N., Harley, A., Thronson, Jr., Capuano, Jr.,
1992, ApJ 393, 611
\bibitem[1994]{worthey}Worthey, G., 1994, ApJS 95,107
\bibitem[1990]{wright}Wright, G. S., James, P. A., Joseph, R. D., McLean, I.
S., 1990, Nature 344, 417
\bibitem[1986]{young1}Young, J. S., Kenney, J., Tacconi, L., Claussen, M.,
Huang, Y-L., Tacconi-Garman, L., Xie, S., Schloerb, F. P., 1986, ApJ 311, L17
\bibitem[1991]{young3}Young, J.S., Scoville, N.Z., 1991, ARA\&A 29, 581
\bibitem[1991]{xu}Xu, C., Sulentic, J.W., 1991, ApJ 374, 407
\bibitem[2001]{erik}Zackrisson, E., Bergvall, N., Olofsson, K., Siebert, A.,
2001, A\&A 375, 814

\end{thebibliography}
\end{document}